\documentclass[aps,prx,showpacs,amsmath,amssymb,superscriptaddress,reprint,10pt,longbibliography]{revtex4-1}

\usepackage{amsmath,amsfonts,amssymb,amsthm,bbm,graphicx,enumerate,times}
\usepackage{mathtools}
\usepackage[usenames,dvipsnames]{color}
\usepackage{comment}
\usepackage{hyperref}
\usepackage{tikz}
\usepackage{todonotes}
\usepackage{graphicx}
\usepackage{textcomp}
\usepackage{float}
\usepackage{siunitx}
\usetikzlibrary{shapes,positioning}
\usepackage{etoolbox}
\newcommand{\propSIlist}[2]{\SIlist[list-units = single,list-final-separator = {, \text{and}~}]{#1}{#2}}

\usepackage{soul}
\setstcolor{red}

	\definecolor{thomas}{rgb}{.0,.4,.0}
  \definecolor{marek}{rgb}{.4,.4,.0}
  \definecolor{alex}{rgb}{.5,.1,.5}
  \definecolor{christian_comments}{rgb}{.75,.1,.5}

\definecolor{bernhard}{rgb}{.8,.0,.0}
\definecolor{bernhard_comments}{rgb}{.58,.0,.82}

\def\p{{\hat \phi}}			
\def\d{{\delta\hat \rho}}	

\def\N{\mathcal N}

\newcommand{\R}{\mathbb{R}}

\renewcommand{\t}{\top}

\renewcommand{\i}{\ensuremath\mathrm{i}}

\newcommand{\id}{1\!\!1}

\newcommand{\expval}[1]{{\langle #1\rangle}}

\def\p{{\hat \phi}}
\def\d{{\delta\hat \rho}}

\def\Vpp{V^{\phi\phi}}
\def\Vpr{V^{\phi\rho}}

\def\Vrr{V^{\rho\rho}}

\renewcommand{\t}{\top}
\def\pp{\hat \varphi}
\def\dd{\delta\hat\varrho}
\def\RD{R_{1\mathrm{D}}}
\def\di{\mathrm{d}}
\def\nGP{n_\mathrm{GP}}
\def\fjd{f_{k}^\rho}
 \def\fjp{f_{k}^\phi}
 \def\gjd{g_{k}^\rho}
 \def\gjp{g_{k}^\phi}

\def\fjk{f_{j,k}^{a,b}}
\def\fkj{f_{k,j}^{a,b}}
\def\NumP{N_p}

\begin{document}
\title{Quantum read-out for cold atomic quantum simulators}

\newcommand\fu{Dahlem Center for Complex Quantum Systems, Freie Universit{\"a}t Berlin, 14195 Berlin, Germany}
\newcommand\tuw{Vienna Center for Quantum Science and Technology, Atominstitut, TU Wien, Stadionallee 2, 1020 Vienna, Austria}
\author{M.\ Gluza}\affiliation{\fu}
\author{T.\ Schweigler}\affiliation{\tuw}
\author{B.\ Rauer}\affiliation{\tuw}
\author{C.\ Krumnow}\affiliation{\fu}
\author{J.\ Schmiedmayer}\affiliation{\tuw}
\author{J.\ Eisert}\affiliation{\fu}
\date{\today}

\begin{abstract}
Quantum simulators allow to explore static and dynamical properties of otherwise intractable quantum many-body systems. 
In many instances, however, it is the read-out that limits such quantum simulations. 
In this work, we introduce a new paradigm of experimental read-out exploiting coherent non-interacting dynamics in order to extract otherwise inaccessible observables. 
Specifically, we present a novel tomographic recovery method allowing to indirectly measure second moments of relative density fluctuations in one-dimensional superfluids which until now eluded direct measurements.
We achieve this by relating second moments of relative phase fluctuations  which are measured at different evolution times through known dynamical equations arising from unitary non-interacting multi-mode dynamics. 
Applying methods from signal processing we reconstruct the full matrix of second moments, including the relative density fluctuations.
We employ the method to investigate equilibrium states, the dynamics of phonon occupation numbers and even to predict recurrences. 
The method opens a new window for quantum simulations with one-dimensional superfluids, enabling a deeper analysis of their equilibration and thermalization dynamics.
\end{abstract}
\maketitle
Quantum simulators offer entirely new perspectives of assessing the intriguing physics of quantum many-body 
systems in and 
out of equilibrium. They are 
experimental setups allowing to probe properties of 
complex quantum systems under unprecedented levels of control \cite{BlochSimulation,CiracZollerSimulation,IonSimulation},
 beyond the possibilities of classical simulations.
Among other platforms, experiments with ultra-cold atoms involving large particle numbers or even continuous quantum fields  have been particularly insightful \cite{Gring_etal12,Hofferberth_etal07,SchmiedmayerGGE,onsolving,Recurrence,StringOrder,Trotzky_etal12,Expansion,BlochMBL,1111.0776,Kaufman}. 

And yet, key questions remain open for a highly unexpected reason: The read-out of state-of-the-art
 quantum simulators is limited. 
 In one-dimensional superfluids, for example, one can probe the dynamics of equilibration \cite{Gring_etal12} occurring in the presence of an effective light-cone  \cite{Hofferberth_etal07} and leading to generalized Gibbs ensembles \cite{SchmiedmayerGGE}.
 The excellent experimental control over that system allowed to observe coherent recurrences in the dynamics of a system of thousands of atoms \cite{Recurrence}.
 However, in that particular setup, further quantifying the recent observations is currently obstructed because only phase quadratures but not canonically conjugate density fluctuations can be measured. 
On the contrary,  if both quadratures could be measured, and hence if genuine \emph{quantum} read-out was possible, then studies of intricate questions on the role of interactions, or entanglement dynamics after a quench could become possible.

This situation is by no means an exception: In fact, in any quantum simulation platform, read-out prescriptions are always restricted in one way or another which constitutes a crucial bottleneck towards studying intricate physical questions. 
For cold atoms in optical lattices, akin to the  development which will be laid out in this work, innovations such as the quantum gas microscope \cite{Sherson-Nature-2010,Bakr-Science-2010,Weitenberg-nature-2011} directly opened up the path towards studying exciting physical phenomena  \cite{StringOrder,Trotzky_etal12,Expansion,BlochMBL,1111.0776,Kaufman}.
Sophisticated read-out methods are therefore  highly desirable and key to a platform.

In this work, we show that quenching a \emph{single} global parameter in the system can  enable a genuine quantum read-out and even allow for state reconstructions.
We hence open a new `window' into a quantum simulator for which so far -- as is common in quantum simulation -- only incoherent, or `classical', read-out was natively possible.

Tomography of many-body systems is typically limited due to the number of necessary observables and complexity of control.
However, often for large systems  the observed dynamics can be described by an effective free field theory capturing the dynamics by means of modes which are long-lived, i.e., not over-damped.
We shall demonstrate that observing their dynamics can already suffice to perform reconstructions of the relevant correlation functions for many-body systems.
The basic principle is that non-equilibrium evolution can mix `quadratures' (non-commuting operators describing the dynamics of a mode) in such a way that consistency of observed correlations implies constraints regarding the unobserved ones.
We will show that they can be quantitatively reconstructed.

Specifically, in this work, we set out to introduce a novel method of tomographic read-out for quantum simulators, by combining known quantum dynamics and available measurements to obtain more information. 
Related ideas of exploiting known random or deterministic unitary dynamics to get access to otherwise inaccessible types of measurements have been theoretically explored \cite{Merkel,Efficient,Renyi,barthel2018fundamental,PhysRevLett.113.045303,ardila2018measuring}. 
However, closest in spirit are tomographic methods in quantum optics where a harmonic rotation in phase space allows to measure two canonically conjugated quadratures by a detector sensitive to only one of them, and hence perform a ``quantum measurement''.

Here, we consider this basic idea in a genuine \emph{multi-mode setting} and demonstrate a \emph{practical} application of the method to one-dimensional \emph{superfluids}.
We acquire data at different times for many modes at the same time and make use of semi-definite programming techniques for achieving reconstructions insusceptible to noise. 
After introducing our new recovery method, we explore the physics of one-dimensional superfluids studying the properties of quench dynamics and its initial conditions.
We use the quantum read-out information concerning both density and phase fluctuations in momentum space to fit the temperature and the global tunnel coupling parameter of the initial state preparation.
Concerning out-of-equilibrium dynamics, we are able to predict recurrences by relying solely on data taken at times away from the recurrence occurrence, demonstrating that the system is coherent throughout the evolution.
Finally, we monitor dynamics of phonon occupation numbers constraining their growth over an extensive observation time giving further experimental evidence for the validity of the effective model.
The successful functioning of our method demonstrates an excellent agreement of the experiment with the theory of  elementary excitations of one-dimensional superfluid  \cite{Cazalilla, MoraCastin}.
Our approach is based on very general and ubiquitous ingredients,  hence the framework that we establish can be expected to be in a natural way applicable to various quantum simulators. 

{\it The system considered.} 
In order to apply our read-out method in practice we will consider the setting of two adjacent 1D Bose gases realized with ultra-cold atoms. 
Their low-energy relative fluctuations in phase and density, $\pp$ and $\dd$, are described by the effective Hamiltonian \cite{MoraCastin, Recurrence}
\begin{equation}
\begin{split}
\hat H=\int_{-\RD}^{\RD} \di z \biggl[&\frac{\hbar^2 \nGP(z)}{4m}\left(\partial_z\pp(z)\right)^2+g(z) \dd(z)^2 \\&+ \frac{\hbar^2 }{4m \nGP(z)}\left(\partial_z\dd(z)\right)^2 \biggr]\ ,\label{eq:H_quench}
\end{split}
\end{equation}
with $m$ being the atomic mass and $\nGP$ the average density profile defined by the ground state of a 1D Gross-Pitaevskii (GP) equation.
The Hamiltonian 
describes phonons which are the elementary density-phase excitations, satisfying bosonic commutation relations $[\dd(z),\pp(z')]=\i\delta(z-z')$. 
The corresponding operators are defined within the atomic cloud whose spatial extension $\RD$ is given by the support of $\nGP$.
Experimentally, one can engineer the density profile $\nGP$ through the trapping potential.
This determines the density-density interaction strength, which is functionally dependent on the density profile,  $g(z) = g[\nGP(z)]$ \cite{Salasnich2002,Recurrence} (see SM).
For typical experimental parameters, the last term in Eq.~\eqref{eq:H_quench} has less importance than the first two \cite{StringariPRL} which together make up the Luttinger model.
In the SM we describe a general numerical scheme for obtaining approximate eigenfunctions of $\hat{H}$  for any $\nGP$ of interest.
For a constant density profile this model gives a linear spectrum and oscillatory eigenfunctions --- which is qualitatively also the case for Eq.\ \eqref{eq:H_quench} even with small GP profile inhomogeneities.

As the excitations are confined within the finite atomic cloud, their spectrum $\{\omega_k, k=1,2,\dots \}$ is discrete \cite{Messiah}.
We denote eigenmode operators of phase and density fluctuations by $\p_k$ and $\d_k$ respectively and use their corresponding wave functions $\fjd,\fjp\in C^2([-\RD,\RD])$ to decompose the real-space fields as
\begin{align}
  \pp(z)&= \sum_{k=0}^\infty \fjp(z) \p_k\ ,\label{eq:decomposition_p}\,\,
  \dd(z)= \sum_{k=0}^\infty \fjd(z) \d_k\ .
\end{align}
Note that here we have $[\d_k,\p_{k'}]=\i\delta_{k,k'}$ where $\delta_{k,k'}$ is the Kronecker symbol in contrast to the Dirac delta in the commutation relations of the real-space fields above.
Written in terms of the eigenmode degrees of freedom $\p_k$ and $\d_k$ the Hamiltonian becomes diagonal
\begin{align}
\hat H=\tfrac12 \sum_{k=1}^\infty \hbar \omega_k (\d_k^2+\p_k^2)+g_0\hat \rho_0^2\ .
\label{eq:H_decoup}
\end{align}
Here $\hat \rho_0$ does not contribute to the visible dynamics because it is conjugate to the global phase which carries no energy and is removed from the data. 
We shall refer to the eigenmode operators as \emph{quadratures} as they constitute a discrete set of bosonic observables which harmonically rotate and do not mix between different modes as can be directly seen from their time evolution within the Heisenberg-picture
\begin{align}
  \p_k(t)=\cos(\omega_k t )\p_k -\sin(\omega_kt)\d_k\ .
  \label{eq:p_t}
\end{align}
In the method presented below, we will make use of approximate eigenmodes obtained from the eigenfunctions of a spatial discretization of the considered model which, for simplicity, we will continue to denote  by $\p_k, \d_k$ and $\fjd,\fjp$ --- see the SM for a more detailed discussion.
Let us stress that by the time evolution in Eq.~\eqref{eq:p_t} density fluctuations, which are not accessible by direct measurements, are dynamically mixed into the observed phase sector which is the foundation to our reconstruction approach.

{\it Quadrature tomography.}
In this section, we turn to describing the  
reconstruction procedure, exploiting the known and efficiently tractable Hamiltonian dynamics
on the one hand and ideas of reconstruction and signal processing on the other. 
This gives rise to a practical and versatile method of reconstructing correlation functions of a type inaccessible
to direct measurement. 
The atom chip experiment \cite{PhysRevLett.84.4749} 
which we are considering here, measures referenced correlation functions of the relative phase through matter-wave interferometry \cite{Schumm05,Recurrence,van2018projective,PhysRevLett113.045303,onsolving,Thomas_thesis} 
\begin{align}
  \Phi( z, z', t) =  \langle (\pp(z,t)-\pp(z_0,t) ) (\pp(z',t )-\pp(z_0,t))\rangle\ .
  \label{eq:obs}
\end{align}
Here we chose to reference the phase with respect to the middle of the system $z_0=\SI{0}{\micro\meter}$ which removes from the data  the global phase canonically conjugate to $\hat \rho_0$.

We aim to reconstruct the second moments of the initial state of the quadratures $\hat r = (\p_1,\ldots ,\p_\N,\d_1,\ldots,\d_\N)^T $ of the $\N$ lowest lying eigenmodes of $\hat{H}$ satisfying  bosonic commutation relations $[\hat r_k, \hat r_{k'}]= \i\Omega_{k,k'}$ with $\Omega=\begin{psmallmatrix}0& \id_\N\\-\id_\N& 0\end{psmallmatrix}$.
For this, we define the \emph{covariance matrix} of the initial state as the collection of second moments
\begin{align}
V_{j,k}=\tfrac12 \expval{\{\hat r_j,\hat r_k\}} = \tfrac12 \expval{ \hat r_j \hat r_k + \hat r_k \hat r_j}\ .
\label{eq:V}
\end{align}
It is important to note that a matrix $V$ constitutes a collection of physically admissible second moments if and only if the matrix $\mathcal Q(V) =  V +\tfrac12\i \Omega \succeq 0$ is positive-semidefinite, i.e., has non-negative eigenvalues.
This condition reflects the Heisenberg uncertainty principle of canonically conjugated observables \cite{PhysRevA.49.1567}.
It will be convenient to use the notation 
\begin{align}
V=\begin{psmallmatrix} \Vpp&\Vpr\\V^{\rho\phi}&\Vrr\end{psmallmatrix}\ .
\end{align}
Note that here we include the interest in both type of correlations, not only those related to phase fluctuations.
Altogether, $V$ allows to provide a \emph{Gaussian} description of the \emph{full} unknown state of the system whose validity can be verified by measuring vanishing higher-order connected correlation functions \cite{onsolving}.
In this work we will consider initial states that are approximately Gaussian and under this assumption determining $V$  yields a full state reconstruction.
For non-Gaussian initial states, the method presented in the following can still reconstruct the second moments $V$, but will not provide specific information on the higher moments.

Using the decomposition into eigenmodes in  Eq.~\eqref{eq:decomposition_p} we obtain for the observable second moments defined in Eq.~\eqref{eq:obs}
\begin{figure}
\includegraphics[width = 0.89\linewidth]{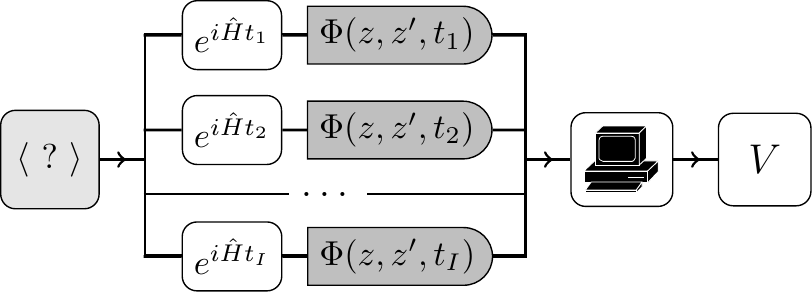}
\caption{\textbf{Recovery procedure.}
How can we measure correlation functions of the elementary excitations in superfluids?
We first note that the Hamiltonian in the system $\hat H$ \eqref{eq:H_quench} is decoupled on the level of momentum space operators $\{\hat\phi_k,\delta\hat \rho_k\}$ involving \emph{multiple} modes rotating at different frequencies \eqref{eq:H_decoup}.
Measurements of real-space \emph{continuum fields} $\{\hat \varphi(z,t)\}$ yield the referenced two-point correlation functions $\Phi(z,z',t)$ defined in Eq.~\eqref{eq:obs}.
In this work, by means of sophisticated post-processing using tools from signal processing, we are able to recover from real-space data taken at equidistant measurement times the full covariance matrix $V$ \eqref{eq:V} of the non-local eigenmodes of the Hamiltonian.
This approach is general and can be applied to any other system in which quenches to non-interacting multi-mode Hamiltonians are available.
}
\label{fig:illustration}
\end{figure}

\begin{align}
  \Phi( z, z', t) = \sum_{j,k=1}^\N f_{z,z'}^{j,k} \Vpp_{j,k}(t)
  \label{eq:obs_rec}
\end{align}
where $f_{z,z'}^{j,k}=(f^\phi_{j}(z)-f^\phi_{j}(z_0) )(f^\phi_{k}(z')-f^\phi_{k}(z_0))$.
Note that we have introduced the cut-off $\N$ in the summation over the eigenmodes, anticipating that higher energy modes will have a negligible contribution in the measured signal either because they carry too much energy to or due to finite real-space resolution in the experiment.

Next, we exploit that the time evolution of the Hamiltonian in Eq.~\eqref{eq:H_decoup}  does not mix quadrature operators of different eigenmodes as stated in Eq.~\eqref{eq:p_t} which gives 
\begin{align}
V(t) = G_t V G_t^T \quad\text{for}\quad 
G_t = \begin{psmallmatrix} 
C_t & - S_t\\
 S_t &C_t
\end{psmallmatrix}
\label{eq:time_ev_cov}
\end{align}
where $C_t = \text{diag}(\cos(\omega_k t ), k=1,\dots, \N)$ and $S_t = \text{diag}(\sin(\omega_k t ) , k=1,\dots, \N)$.

As summarized in Fig.~\ref{fig:illustration}, we now have all ingredients needed for quantum read-out.
Specifically, based on the  relations \eqref{eq:obs_rec} and \eqref{eq:time_ev_cov}, we can 
recover the density correlations through a least squares recovery problem.
For this we collect all measured values of $\Phi( z, z', t_i)$ at different points $z$ and $z'$ and times $t_i$ in a vector $b$.
Furthermore we define a linear map $\mathcal{A}(\tilde V)$ which, given some trial covariance matrix $\tilde V$, outputs the values of $\Phi( z, z', t_i)$ sorted as in  $b$ via Eq.~\eqref{eq:obs_rec}.
The time-evolution is implemented using Eq.~\eqref{eq:time_ev_cov} such that only the covariance matrix of the initial state is used to fit the observed data.
If $W$ denotes a weighting matrix then $W(\mathcal A(\tilde V)-b)$ is the vector of the weighted least squares residues.
The covariance matrix optimally fitting the data is then given by the solution to the following optimization problem
\begin{align}
\begin{split}
	\Theta~ =&\min_{\tilde V}~\|W \mathcal A(\tilde V) - W b\|_2,\\
	\text{subject to} &\quad  \mathcal Q(\tilde V) = \tilde V +\tfrac12\i \Omega \succeq 0\ .
\end{split}
\label{eq:LS}
\end{align}
The first line implements the minimization of the length $\|\cdot\|_2$ of the vector of weighted least squares residues and the condition in the second line ensures that $V$ is a physical covariance matrix.
The optimal solution to this  convex quadratic problem with a semi-definite constraint yields the covariance matrix $V$ of the initial state with a minimal value of $\Theta$.
The optimization can be performed  efficiently and reliably numerically with standard methods for semi-definite programming. 
We use the package \texttt{cvx} \cite{cvx}, see the SM for more details on the implementation.
The most basic idea of an algorithm solving \eqref{eq:LS} is to repeatedly take a steepest-descent step towards minimizing the least squares residue and impose $\mathcal Q(V)  \succeq 0$. 
Standard convex optimization packages like \texttt{cvx} solve such a problem in a more sophisticated way ensuring numerical accuracy and converge in a matter of seconds.
In the implementation we chose a diagonal weighting matrix $W$ with entries $\sigma[\Phi(z,z',t_i)]^{-1}$ where $\sigma[\Phi(z,z',t_i)]$ denotes the standard deviation of each measured value.
This weighting allows to put more emphasis on more precise values and yields with this a more reliable scheme as we find that $\Phi(z,z',t)$ grows typically for increasing spatial separations $|z-z'|$ but $\Phi(z,z',t)/\sigma[\Phi(z,z',t)] \approx\texttt{const}$.

Note that the above idea and in fact the whole framework of quantum read-out formulated here is independent of the dimensionality of the Hamiltonian and can be applied, e.g., in two dimensions.
There is also no restriction to continuum systems so lattice models can be treated similarly.

\begin{figure*}
\includegraphics[width=1\linewidth]{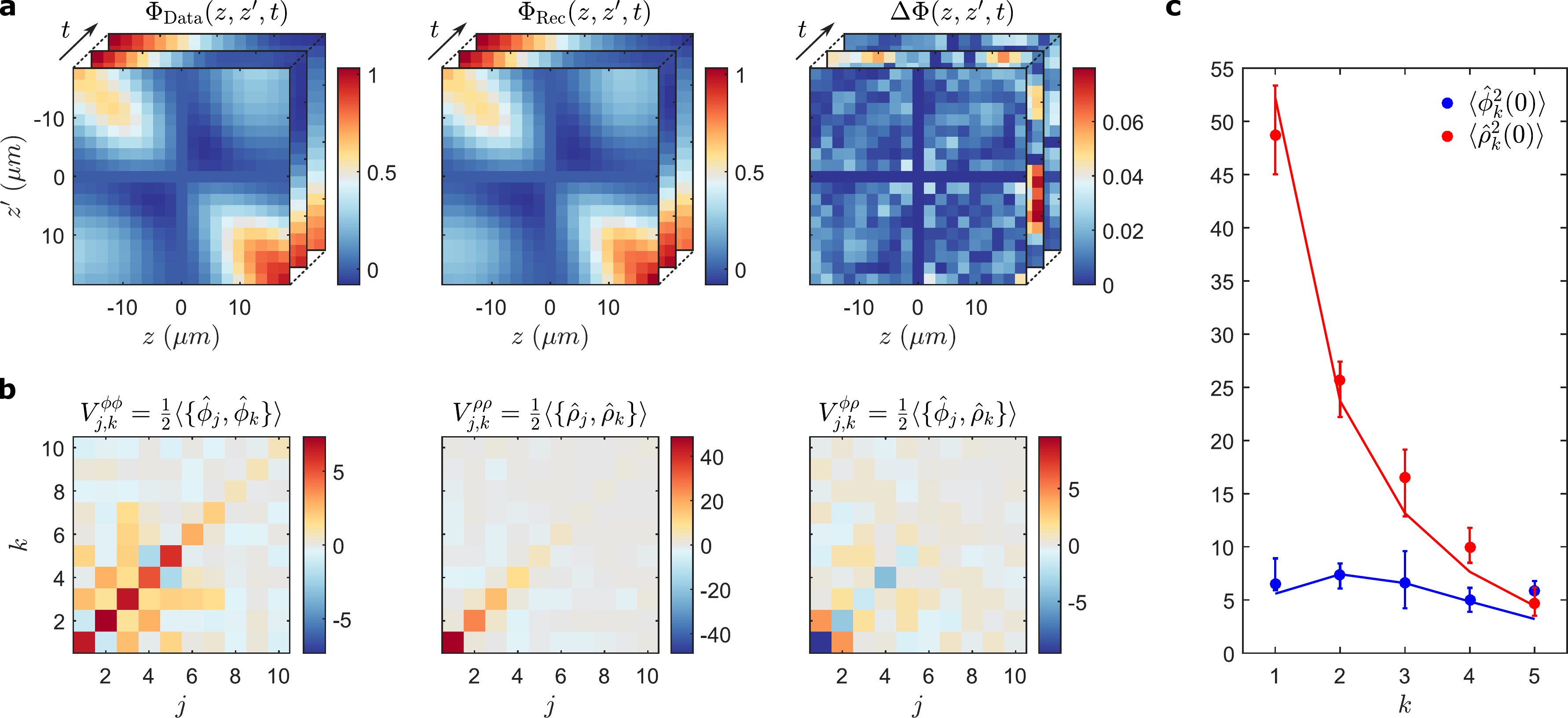}

\caption{
\textbf{Initial state reconstruction.} 
Reconstruction of the full initial state right after decoupling at $t=\SI{0}{\milli\second}$ based on the phase correlations measured during the dephasing dynamics immediately after the quench ($t=\propSIlist{1;3.5;6;8.5;11;13.5}{\milli\second}$).
\textbf{(a)}
Comparison of the measured phase correlations $\Phi_\mathrm{Data}(z,z',t)$ (left) to the reconstructed ones $\Phi_\mathrm{Rec}$ (center).
The time slice presented in the foreground corresponds to $t=\SI{1}{\ms}$.
We find that the reconstruction yields good agreement with the data, evidenced by the respective weighted difference to the data
$\Delta\Phi = |\Phi_\mathrm{Data}-\Phi_\mathrm{Rec}|/\sigma[\Phi_\mathrm{Data}]$ (right). 
\textbf{(b)}
Reconstructed covariance matrix $V$ of the initial state at $t=0$ for the eigenmodes $j,k=1,\ldots,10$.
From left to right the reconstructed phase-phase $\Vpp$, density-density $\Vrr$ and phase-density correlations $\Vpr$ are plotted. 
The correlations $\Vpp$ and $\Vrr$ are close to diagonal in the numerically obtained wave functions $\fjp$, indicating that the eigenmodes of the system are well captured.
For an initial thermal state of the Hamiltonian \eqref{eq:H_ini} the cross-correlations should vanish $\Vpr \equiv 0$, here we find a small contribution.
Note that the influence of higher-energy modes is suppressed by the limited spatial resolution in the experiment (see SM).
\textbf{(c)}
Comparison of the diagonal elements of $\Vpp$ (blue bullets) and $\Vrr$ (red bullets) at $t=0$ with the predictions for a thermal state of the pre-quench Hamiltonian given in \eqref{eq:H_ini} (solid lines). The error bars correspond to the $80\%$ confidence intervals obtained from a bootstrap analysis \cite{efron1986bootstrap}.
The thermal predictions are corrected for the suppression due to the finite imaging resolution, see SM and Ref.~\cite{Thomas_thesis}.
We find the correlations of the first five modes to agree well with a thermal state at $T=\SI{52}{\nano\kelvin}$ and $J = 2\pi \times \SI{1.1}{\hertz}$ obtained by a combined least squares fit of $\langle\p_k^2\rangle$ and $\langle \d_k^2\rangle$.
The strong suppression of the higher mode signals by the imaging renders a meaningful comparison impossible.
}
\label{fig:V}
\end{figure*}

{\it Experimental data analysis.}
Let us consider the state preparation procedure used in the recent experiment \cite{Recurrence} where recurrent dynamics has been observed.
Given an estimate of the average number of atoms per gas  $N_{\rm Avg} \simeq 3400$ and the shape of the experimental box-like potential we can numerically obtain the average density profile $\nGP(z)$ from the GP equation.
This specifies the Hamiltonian  $\hat H$ \eqref{eq:H_quench} with $\RD \simeq \SI{25}{\micro\meter}$.
Hence, we have the full information necessary to compute the eigenmode wave functions needed for the reconstruction procedure in Eq.~\eqref{eq:obs_rec}.
The number of relevant  wave functions $\N\approx 10$ can be upper bounded by considering the finite resolution of the interference images.
In our case the phase fluctuations $\Phi(z,z',t)$ defined in Eq.~\eqref{eq:obs} can be measured at points $z,z'$ spaced by the pixel size of the camera $\delta \approx \SI{2}{\micro\meter}$ \cite{Thomas_thesis}. 
In addition, other effects including diffraction limit the resolution.
The measured values can be related to theoretical continuum predictions by implementing a real-space cut-off via a Gaussian convolution with standard-deviation $\sigma\approx \SI{3.5}{\micro\meter}$ (see SM).

Initially the two adjacent gases whose relative phase fluctuations we are studying are strongly coupled.
Hence, the state preparation is to a good approximation governed by 
\begin{align}
\hat H_\text{Ini} = \hat H + J \int_{-\RD}^{\RD} \di z \ \nGP(z) \pp(z)^2
\label{eq:H_ini}
\end{align}
where the tunnel coupling term of strength $J$ is pulling the relative phase field to zero (i.e., $\langle\mathrm{cos}(\pp)\rangle \simeq 1$).
The initial state can be expected to be a low temperature thermal state of $\hat H_\text{Ini}$.
Following that preparation, the system is quenched by suddenly turning off the tunnel coupling.
Experimentally, this is realized by separating the two gases over a time of \SI{2}{\milli \second} until $J$ drops to zero.
The middle of this ramp defines the initial time $t=\SI{0}{\milli\second}$ and the subsequent evolution under the Hamiltonian $\hat H$ \eqref{eq:H_quench} is measured in steps $\Delta t = \SI{2.5}{\milli \second}$.

Based on the data from this initial dynamics we can reconstruct the initial state at $t=\SI{0}{\milli\second}$.
In Fig.~\ref{fig:V}a we plot the reconstructed phase correlations, showing good agreement with the measured values signifying the consistency of our method.
The corresponding covariance matrix of the full initial state is shown in Fig.~\ref{fig:V}b.
Most importantly, note that we are indeed able to infer density fluctuations of the form $\langle \d_j\d_k\rangle$.
Using the mode transformation from Eq.~\eqref{eq:decomposition_p}, this information from the eigenmode-space can be also translated to real-space.
However, many physical properties of the initial state can be directly extracted from the eigenmode correlations.
We firstly observe that the blocks $\Vpp$ and $\Vrr$ are close to being diagonal.
Hence, we find that the collective modes of the system are well captured by the numerically obtained wave functions $\fjp$.
This supports the expectation that the initial state is thermal with respect to the pre-quench Hamiltonian~\eqref{eq:H_ini}.
As the eigenmode wave functions are not strongly affected by the quench for the chosen trap geometry (see SM), if the system was thermal with respect to the initial Hamiltonian,  the reconstructed state should remain diagonal even when expressed in terms of the wave functions of the quench Hamiltonian.

\begin{figure*}
\includegraphics[width=0.8\linewidth]{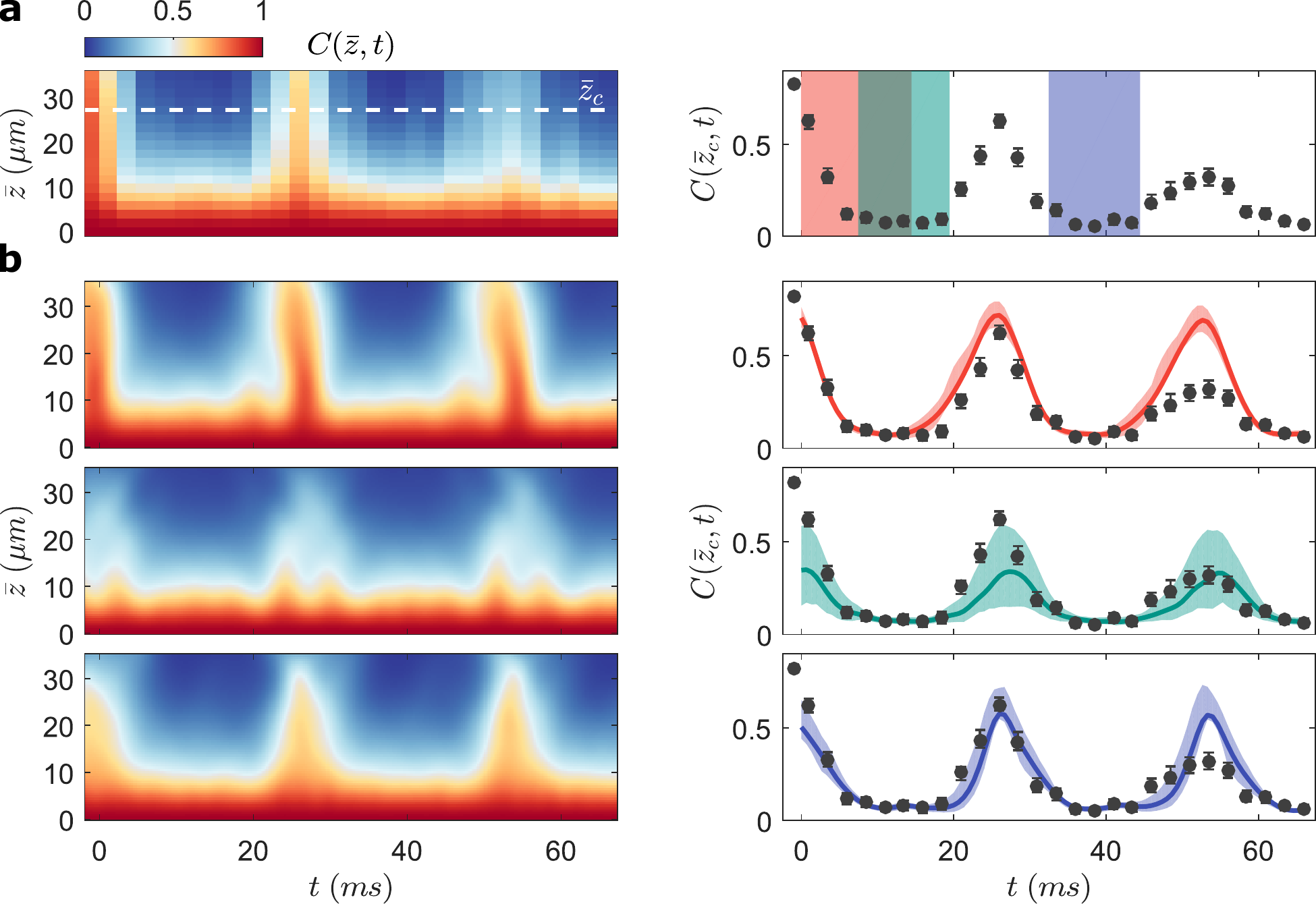}
\caption{\textbf{Dynamical predictions.}
\textbf{(a)}
Measured dynamical phase correlations $C(\bar{z}, t)$ (left) shown together with a  cut at $\bar{z}_c = \SI{27.25}{\micro\meter}$ (right).
The colored intervals indicate the time window used in the reconstructions presented in (b) (red: upper row, green: middle row, blue: lower row).
\textbf{(b)}
Following reconstructions based on input data from a given time window we calculated the phase correlation functions  $C(\bar{z}, t)$ capturing the full spatio-temporal dynamics (left).
The cut at $\bar{z}_c$  (right) shows the quantitative comparison to the measured data.
The shaded area around the curves, as well as the error bars of the data, indicate $80\%$ confidence intervals obtained from a bootstrap analysis \cite{efron1986bootstrap}.
The upper row corresponds to the propagation of the initial state reconstruction presented in Fig.~\ref{fig:V} (red interval).
The middle and lower row show the propagation of reconstructions based on seemingly dephased data (green and blue interval, respectively).
While the dynamical prediction based on the initial dephasing dynamics works well the reconstructions based on data taken in between the recurrences is limited by the finite sample size (green interval) -- an effect reproduced by numerical simulations (see SM). 
Note that the quantitative discrepancy at times far away from input intervals is due to terms of higher order not captured by the effective theory of Eq.~\eqref{eq:H_quench}.
}
\label{fig:revivals}
\end{figure*}

On the other hand, we remark that allowing for off-diagonal correlations and cross correlations in $V$ is necessary for an accurate reconstruction as otherwise the comparison in  Fig.~\ref{fig:V}a is significantly worse.
One reason for their presence can be small deviations between the assumed eigenmodes and the true eigenmodes of the system, e.g., due to  pre- or post-quench trapping potential imperfections not included in the GP profile.
Another reason could be that the  initial state is genuinely out of thermal equilibrium which would be interesting from the quantum information perspective in the context of the resource theory of coherence \cite{coherence_gaussian, coherence_review}.

Independent of these subtleties, our read-out method allows us to study how the energy is distributed in the system based on the measured out-of-equilibrium phase fluctuations.
We now have access to the phonon occupation numbers given by $ n_k=\tfrac 12  \langle \p_k^2+\d_k^2\rangle-\tfrac 12$  and can study energy expectation values $\langle \hat H\rangle = \sum_{k=1}^\N \omega_k n_k$ with $\hat H$ given in Eq.~\eqref{eq:H_quench}.
More specifically, we can check from the observed data if the energy is distributed among the modes in a thermal way.
The gas is prepared in a double well trap with large tunnel coupling and we expect the prepared state to be thermal with respect to the Hamiltonian \eqref{eq:H_ini}.
Based on Fig.~\ref{fig:V}b we find that the reconstructed initial state shows significant suppression of the phase fluctuations which are an order of magnitude smaller than the density fluctuations.
This is consistent with the initial energetic penalty on phase fluctuations.
In Fig.~\ref{fig:V}c we show the quantitative comparison of the reconstructed second moments $\langle\p^2_k\rangle$ and $\langle\d^2_k\rangle$ compared to the thermal theory with the Hamiltonian $\eqref{eq:H_ini}$.
Due to the finite imaging resolution we are only able to resolve the lowest-lying eigenmodes. 
We find that their second moments agree with a thermal distribution of the coupled Hamiltonian  \eqref{eq:H_ini}.
On the other hand, we also observed that the reconstructed initial state does not agree with the experimental state before the start of the decoupling ramp, as it leads to weaker phase locking than what was measured.
This hints at the finite decoupling ramp having a significant influence on the correlations observed in the quench dynamics despite the tunnel coupling decreasing exponentially in the height of the barrier that is being ramped-up.
The reconstruction hence extracts an effective initial state of the dynamics.
If the physics of the initial Hamiltonian is of particular interest, then this effect might be diminished by performing a faster quench to the free system.
On the other hand, let us remark that the physics of quenches of this type is key to the observation of generalised Gibbs ensembles in the considered setup \cite{SchmiedmayerGGE}.
In general it is difficult to model the complete process of state preparation theoretically as it involves a strongly correlated phase of the sine-Gordon model out of equilibrium \cite{onsolving} and our method could offer new experimental insights.

\begin{figure*}
\includegraphics[width=0.9\linewidth ]{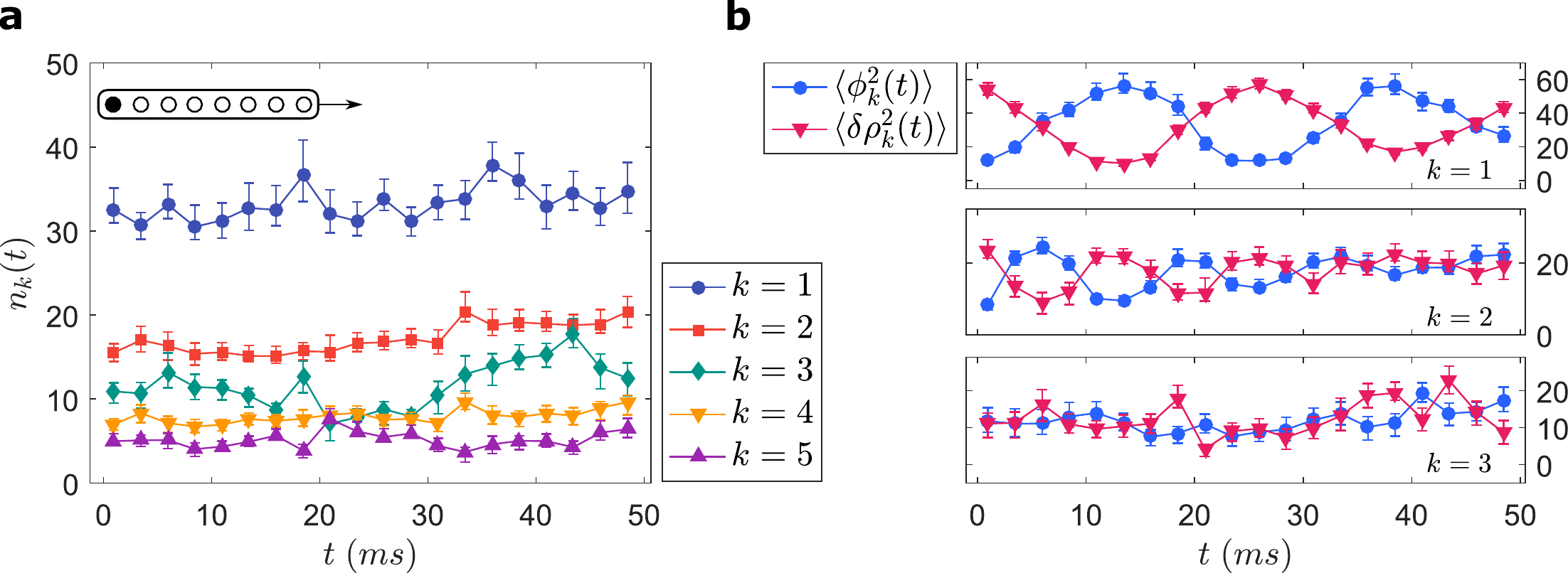}
\caption{\textbf{Phonon occupation dynamics.} 
\textbf{(a)}
Phonon occupation numbers $n_k(t)$ of the first five 
modes $k=1,\dots, 5$ as a function of time $t$.
Each point is based on a reconstruction with an input interval $\{t, t+\Delta t, \ldots, t+7\Delta t\}$ of length $I=8$, illustrated by the black box in the upper left corner. 
For the ideal mean-field model $n_k$ should be a constant of motion.
\textbf{(b)}
Time-resolved central moments of the phase and density fluctuations in momentum space $\langle \p_k^2(t)\rangle$ and $\langle \d_k^2(t)\rangle$ for the first three modes $k=1,2,3$ (top to bottom).
We observe a gradual decay of the oscillation amplitude reflecting the apparent equilibration observed in Fig.\ \ref{fig:revivals}.
The error bars indicate the $80\%$ confidence intervals obtained from a bootstrap analysis \cite{efron1986bootstrap}.
The lines connecting the data points are a guide to the eye.
}
\label{fig:occupations}
\end{figure*}
{\it Recurrent dynamics.}
With the reconstruction of the full state of the system also its evolution beyond the interval of input times can be calculated.
Propagating the covariance matrix $V$ forward or backward in time via \eqref{eq:time_ev_cov} allows us to pre- and redict the system's dynamics. 
In Ref.~\cite{Recurrence} this dynamics was visualized and quantified through the correlator
\begin{equation}
C( |z-z'|,t ) = \left\langle \cos\bigl(\pp(z,t)-\pp(z',t) \bigr)\right\rangle \ .
\label{eq:def_cal_c}
\end{equation}
The phase-locked initial state corresponds to $C \approx 1$ independent of the longitudinal separation $\bar{z} = |z-z'|$.
In Fig.~\ref{fig:revivals}a, we show $C(\bar{z},t)$ obtained from the experimental data.
Due to a linear dispersion relation and an equally spaced spectrum, the involved modes start to rephase after the inital dephasing dynamics leading to partial recurrences of the initial state \cite{Recurrence}. 
Fig.~\ref{fig:revivals}b shows how this rephasing dynamics can be predicted from the reconstructed states.
For the reconstruction based on the initial dephasing dynamics, for example, we obtain a good qualitative prediction of the recurrences (red interval).
Quantitative agreement is lost over time due to interaction effects between the modes.
These interactions are mediated by higher-order terms beyond the effective model assumed in \eqref{eq:H_quench} and can therefore not be captured \cite{MoraCastin}.
Nevertheless, the reconstruction method is robust enough such that we can obtain an accurate short-time prediction even using data that is seemingly fully dephased, i.e., where $C(\bar{z},t)$ is nearly indistinguishable from the correlations of a thermal state of the quench Hamiltonian $\hat H$ (blue interval).
However, in some cases (green interval) we find that statistical fluctuations can lead to large error bars, an  effect reproduced by numerical simulations (see SM).
Note also that the last two intervals were intentionally chosen to be short, including only seemingly dephased data between the recurrences.
They cover $I=5$ input times, during which the slowest eigenmode performs only about a quarter of a rotation.
Therefore, the influence of the finite statistical sample size is more severe in these reconstructions.

{\it Phonon occupation dynamics.}
Besides providing new insights into the state preparation, access to the full covariance matrix can enable entirely new ways of exploring the effects of interactions.
The effective model given in \eqref{eq:H_quench} is obtained in a perturbative expansion of the Lieb-Liniger Hamiltonian up to second order \cite{MoraCastin}.
For long evolution times, however, the dynamics  can also be affected by the neglected terms that, e.g., can give rise to effects such as Beliaev-Landau damping \cite{MoraCastin}.
It is challenging to obtain the rates of such processes by numerical calculations as interacting bosonic dynamics are notoriously difficult to treat  and  various approximations are necessary \cite{mazets2018integer, POLKOVNIKOV20101790, huber2018thermalization,kukuljan2018correlation}.
Therefore, it would be interesting to use the atom chip experiments to measure the damping rates and compare with theoretical predictions to validate different methods.

Here we show how the recovery method described above can be used to investigate these higher-order processes.
To that end, we perform the recovery procedure for different input intervals of length $I=8$, with varying starting points.
For each interval we obtain an estimate of the central moments of phase and density fluctuations, $\langle \p_k^2(t)\rangle$ and $\langle \d_k^2(t)\rangle$, and calculate the phonon occupation numbers $ n_k(t)=\tfrac 12  \langle \p_k^2(t)+\d_k^2(t)\rangle-\tfrac 12$. 
Scanning the starting point of the input interval through the measurement times allows us to investigate the dynamics of these observables, as shown in Fig.~\ref{fig:occupations}.
The interval length is chosen long enough such that the slowest eigenmode picks up enough dynamical phase to ensure a stable reconstruction.
At the same time, it is chosen short enough such that interactions between the modes do not  influence the reconstruction.  

The occupation numbers $n_k$ are constants of motion of $\hat{H}$.
In Fig.~\ref{fig:occupations}a, we show their reconstructed dynamics  for the five lowest eigenmodes.
We find that overall the occupation numbers do not vary strongly and stay almost constant.
This is expected as perturbations to the quench Hamiltonian should be negligible, and in any case they are irrelevant in the sense of the renormalization group.
Note, however, that for different measurements with other system sizes we find indications of a trend of slowly increasing mode occupations (see SM).
While the dynamics of occupation numbers is constant, Fig.~\ref{fig:occupations}b shows how at the same time the individual modes rotate between phase and density fluctuations.
We find that this dynamics is clearly damped.
This hints that the source of the recurrence damping observed in Fig.~\ref{fig:revivals}a and Ref. \cite{Recurrence} is a loss of the initial quadrature `squeezing'  $\langle \p_k^2(0)\rangle / \langle \d_k^2(0)\rangle \ll 1$ within each mode $k$ rather than changes in their occupations.

Our method makes it possible to extract mode resolved damping rates of the collective excitations: In the future using smaller time steps $\Delta t$  and possibly non-equidistant measurement times should allow to study also higher modes and test theoretical predictions concerning the dynamics under perturbations to the non-interacting effective model.

{\it Discussion and outlook.}
We have formulated and demonstrated the functioning of a new quantum read-out method for quantum simulators where we reconstruct the second moments of pairs of conjugated observables by measuring at different times only one of them.
The developed scheme allows us to reliably reconstruct the covariance matrix of non-local low-energy excitations of a one-dimensional superfluid based on experimental data from an atom chip experiment which makes phase measurements but does not directly access density fluctuations.

We found several interesting insights into the physics of the system.
Firstly, the strong energetic penalty on phase fluctuations present during the state preparation is reflected in the reconstructed correlations as there are significantly less phase than density fluctuations.
The reconstructed state is almost diagonal which underlines that the eigenmodes before and after the quench are closely related and on a higher level demonstrates that the effective theory captures the relevant degrees of freedom of the system.
A fit to a thermal model for the initial state allowed us to estimate the temparature and the effective tunnel coupling in the state preparation.
In the considered setting, recurrences of the system have been recently observed \cite{Recurrence} which are due to an approximately linear spectrum of the phonons.
We have demonstrated that our method can take input data from times when the system is seemingly dephased in order to predict recurrences by evolving the reconstructed covariance matrix in time, strongly underlining the predictive power of the obtained recovery scheme.
Finally, we have studied the occupation numbers of the eigenmodes over time and obtained strong constraints on the rate of their growth.
We have reconstructed the contribution of phase and density fluctuations over time and found that their oscillations are damped.
We expect that a quantitative experimental assessment of possible reasons of the deviations from the non-interacting effective model will become possible by following the lines of this work.

Our work paves the way towards new intriguing experiments by giving access to quadrature operators which can be used as the basic ingredients for many quantum information processing protocols \cite{GaussianQuantumInfo,Continuous, schnabel2017squeezed}. 
The method presented offers a novel window into quantum simulators, allowing to assess initial states, notions of entanglement and various other quantities previous read-out schemes did not allow for.
It is our hope that our new quantum read-out method will enable exciting insights into the physics of ultra-cold superfluids, but also due to its generality that it will become a versatile tool used in state-of-the art quantum technologies allowing to fully use the power of the existing quantum simulation platforms \cite{Roadmap}.

{\it Note added:} After the completion of the manuscript, we became aware of similar developments in the discrete setting of optical lattices with applications to topological band insulators \cite{PhysRevLett113.045303,ardila2018measuring,tarnowski2017characterizing} -- it would be interesting to also include there our ideas of using semi-definite constraints ensuring that the 
reconstructed covariance matrix is physical and the recovery stable.

{\it Acknowledgements.} 
We thank C. Riofrio, F.\ Essler, I.\ Mazets, A.\ Steffens for useful discussions and comments. This work has been supported by the ERC (TAQ, QuantumRelax), the 
European Commission (AQuS), 
the German DFG (FOR 2724, CRC 183, EI 519/14-1, 519/9-1, EI 519/7-1), the Templeton Foundation,
the Austrian Science Fund (FWF) through the doctoral program CoQuS (W1210) (T.S., B.R.) 
and the SFB 1225 ‘ISOQUANT’ financed by the DFG and the FWF.
This work has also received funding from the European Union's Horizon 2020
research and innovation programme under grant agreement No.~817482 (PASQuanS).

%

\appendix
\begin{widetext}
\tableofcontents

This supplemental material is structured as follows.
In Sec.~\ref{sec:app:eigenmodes} begin by explaining how to obtain eigenmodes in which the dynamics is decoupled and is a simple rotation of the eigenmodes.
In Sec.~\ref{sec:app:recoverydetails} we give more details concerning the reconstruction procedure and its implementation.
In Sec.~\ref{sec:app:simulateddata} we show the functioning of the recovery procedure on simulated Gaussian data.
Finally, in Sec.~\ref{sec:app:extendeddata} we show figures analogous to the main text but based on additional data for systems of different sizes.
\section{Calculating the eigenmodes}
\label{sec:app:eigenmodes} 
In this section we describe in detail how we obtain the eigenmodes of the quench Hamiltonian which can be viewed as a CFT in curved space-time background whenever the GP profile is not homogeneous.
In this case, a discretization of fields allows to approximate the low-lying eigenmodes of the continuum Hamiltonian by the eigenmodes of a Hamiltonian involving a finite number of degrees of freedom which are the average fields in a given discretization cell.
We then show how to diagonalize the coarse-grained Hamiltonian numerically taking into account that the quench Hamiltonian has a zero-mode.
Finally, we describe how to use the numerically obtained wavefunctions for finitely many modes as an approximation to the corresponding eigenmodes in the continuum limit.

The Hamiltonian describing the quench dynamics is functionally parametrized by the GP profile $\nGP$.
Due to transverse broadening of the wave functions \cite{Recurrence} the density-density interaction is functionally dependent on the GP profile and reads 
\begin{align}
g(z) =\hbar \omega_\perp a_s {(2 + 3 a_s \nGP(z))}/{ (1+2a_s\nGP(z))^{3/2}}
\end{align}
where $\omega_\perp$ is the radial trapping frequency and  $a_s$ is the scattering length \cite{Salasnich2002}.
Hence, by knowing the GP profile, we know the Hamiltonian and so we can find the eigenmodes.
Here we show how to do this even if the GP profile is not homogeneous $\nGP\neq \texttt{const}$
\subsection{Discretization of fields}
\newcommand{\discq}{\hat{Q}}
\newcommand{\discpp}{\pp^{(N)}}
\newcommand{\discdd}{\dd^{(N)}}
\newcommand{\discp}{\p^{(N)}}
\newcommand{\discd}{\d^{(N)}}
We want to find approximations to the wave functions and eigenmodes discussed above by discretizing the interval $[-\RD,\RD]$ into $N$ pixels, each of size $2\RD/N$.
Fixing $N$, for $l=1,\ldots, N+1$ the coordinates of the discretization lattice read $z_l = -\RD + 2 \RD \tfrac{l-1} {N}$ and we define discretization pixels which are the closed intervals $p_l=[z_l,z_{l+1}]$ for $l=1,\dots,N$.
We then introduce the discretized operators as the integration of the field operators via
\begin{align}
 \discpp_l =  {\frac{ 1 }{\Delta z} }\int_{p_l} \di z\ \pp( z),
\label{eq:x_discrete1}\\
\discdd_{l} = {\frac{ 1 }{\Delta z} }\int_{p_l} \di z\  \dd( z),
\label{eq:x_discrete2}
\end{align}
with $\Delta z := |p_l|= {2 \RD}/{N}$. 
Following Refs.\ \cite{GaussianQuantumInfo,Continuous}, these discretized operators yields a vector of canonical coordinates 
\begin{align}
\discq =(\discpp_1,\ldots \discpp_N,\discdd_1\ldots\discdd_N)^T,
\end{align}
satisfying the bosonic canonical commutation relations  $[\discq_j, \discq_k]=\i\Omega_{j,k}/\Delta z$ where $\Omega=\begin{psmallmatrix}0& \id_N\\-\id_N& 0\end{psmallmatrix}$.
as can be verified easily.
Observe that the right-hand side will yield a Dirac delta in the continuum limit $N\rightarrow\infty$.
The discretization of the effective model will be a quadratic operator in the discretized modes $\discpp_l$ and $\discdd_l$ which can be efficiently diagonalized using single particle transformations only as we want to explain below in the next section. 

\subsection{Decoupling of the effective model using symplectic transformations}
Using the general notation of quadratures $\discq$, we consider quadratic Hamiltonians of the form
\begin{align}
  \hat H_N=\tfrac12 \discq^T H \discq = \tfrac12 \sum_{j,k=1}^{2N}H_{j,k}\discq_j\discq_k ,
  \label{eq:app:H}
\end{align}
where $H=H^\t\in\mathbb R^{2N\times 2N}$ are the couplings
We will assume that $H$ is positive semi-definite, i.e., $H\succeq 0$ and that there is no coupling between the phases and densities in the effective model and all Hamiltonians considered in this work will have this property.
In this case the couplings $H$ will be block diagonal and we will use the decomposition 
\begin{equation}
H=\begin{pmatrix}H_\phi& 0\\0& H_\rho\end{pmatrix}= H_\phi\oplus H_\rho\ .
\end{equation}

To discretize the integral we define the geometric mean $\eta_l = ({ \nGP(z_l) \nGP(z_{l+1})})^{1/2}$ for $l=1\dots,N$ which gives
\begin{align}
\hat H&\approx \Delta z \sum_{l=1}^{N-1} \frac{\hbar^2 \eta_l }{4m}\left(\frac{ \discpp_l-\discpp_{l+1}}{\Delta z}\right)^2 +\Delta z \sum_{l=1}^N g(z_l) \discdd_l{}^2 + \Delta z \sum_{l=1}^{N-1} \frac{\hbar^2}{4m \eta_l}\left(\frac{ \discdd_l-\discdd_{l+1}}{\Delta z}\right)^2 \\
&=\Delta z \sum_{l=1}^{N-1}\left[\frac{\hbar^2 \eta_l}{4m}\left(\frac{\discq_l-\discq_{l+1}}{\Delta z}\right)^2\right]+ \Delta z \sum_{l=1}^N g(z_l) \discq_{l+N}^2 + \Delta z \sum_{l=1}^{N-1}\left[\frac{\hbar^2 }{4m\eta_l}\left(\frac{\discq_{l+N}-\discq_{l+1+N}}{\Delta z}\right)^2\right]\\
  &:=\hat H_N
\ .
\end{align}
From this we read off
\begin{align}
  H_\phi &= \frac{\hbar^2 }{2m \Delta z } 
  \begin{pmatrix} 
  \eta_1 &-\eta_1\\
  -\eta_1 &\eta_1+\eta_2 &-\eta_2\\
  &&\ddots\\
 &  & -\eta_{N-2} &\eta_{N-2}+\eta_{N-1} &-\eta_{N-1}\\
& && -\eta_{N-1} &\eta_{N-1}
  \end{pmatrix},\\
  H_\rho &= 2 \Delta z 
  \begin{pmatrix} g(z_1)\\&g(z_2)\\&&\ddots\\&&&g(z_N)\ 
  \end{pmatrix}
  +
  \frac{\hbar^2 }{2m \Delta z } 
  \begin{pmatrix} 
  \eta_1^{-1} &-\eta_1^{-1}\\
  -\eta_1^{-1} &\eta_1^{-1}+\eta_2^{-1} &-\eta_2^{-1}\\
  &&\ddots\\
 &  & -\eta_{N-2}^{-1} &\eta^{-1}_{N-2}+\eta^{-1}_{N-1} &-\eta^{-1}_{N-1}\\
& && -\eta^{-1}_{N-1} &\eta^{-1}_{N-1}
  \end{pmatrix}
 \ .
\end{align}
Depending on the detail of the simulation $g(z)\approx g(0)$ can be taken constant or $H_\rho$ may include the pressure term as discussed above with a similar discretization scheme.
With this notation, we obtain 
\begin{align}
  \hat H_N = \tfrac12 \hat x^\t ( H_\phi\oplus H_\rho ) \hat x \approx\hat H\ .
\end{align}

Starting from a set of canonical coordinates $\discq$ then $\hat{r} = S\discq$ for $S\in\R^{2N\times 2N}$ will again denote a vector of canonically commuting operators if $S$ is symplectic, i.e., it fulfills
\begin{equation}
 S\, \Omega\, S^T = \Omega \end{equation}
which can be seen by explicitly checking that $\hat{r}$ again fulfills  $[\hat{r}_j,\hat{r}_k ] = \i \Omega_{j,k} / \Delta z$.

In view of diagonalizing the Hamiltonians of interest, it is important to note that matrices of the form $S = Q\oplus Q$ for any orthogonal $Q\in O(N)$ as well as $S = A \oplus A^{-1}$ for any invertible $A\in GL(N,\R)$ that is symmetric, i.e., $A^T=A$ are both symplectic matrices and that the inverse as well as the product of symplectic matrices are again symplectic. 
We can then diagonalize Hamiltonians of the form as given in Eq.~\eqref{eq:app:H} under the assumption that $H_\rho$ is invertible.
This property allows us to define a symplectic matrix
\begin{equation}
 S_1 = \begin{pmatrix}(H_\rho)^{1/2}&0\\ 0&(H_\rho)^{-1/2}\end{pmatrix}
 \end{equation}
such that 
\begin{equation}
 S_1^{T} H S_1 = ((H_\rho^T)^{1/2} H_\phi (H_\rho)^{1/2})\oplus \id_N\ .
\end{equation}
 The matrix of the phase couplings in the new coordinates reads $\tilde H_\phi = (H_\rho^T)^{1/2} H_\phi (H_\rho)^{1/2} $ and is again real and symmetric such that it can be diagonalized by an orthogonal transformation $Q\in O(N)$ with $\tilde H_\phi= Q\Sigma Q^T$. 
 Here, $\Sigma$ is diagonal and we assume that all zero eigenvalues are sorted to the first $N_0 \geq0$ positions, i.e., $\Sigma = 0_{N_0}\oplus \tilde{\Sigma}$ with $\tilde{\Sigma}\succ0$ diagonal and we define the eigenfrequencies $\omega$ via $\tilde{\Sigma}^{1/2} = \mathrm{diag}(\omega_{N_0+1},\dots,\omega_N)$.
 With the diagonal matrix  $\Sigma_\phi = \id_{N_0} \oplus \tilde{\Sigma}$ and the transformation
 \begin{equation}
  S_2 = \begin{pmatrix}Q_\phi&0\\ 0&Q_\phi\end{pmatrix}\begin{pmatrix}\Sigma_\phi^{-1/4}&0\\0& \Sigma_\phi^{1/4}\end{pmatrix}
 \end{equation}
 we obtain
 \begin{equation}
  S_2^TS_1^T H S_1 S_2 = (0_{N_0} \oplus \tilde \Sigma^{1/2})\oplus (\id_{N_0} \oplus \tilde \Sigma^{1/2}) \ .
 \end{equation}
That is, in the canonical coordinates $(\discp_1,\ldots \discp_N,\discd_1\ldots\discd_N)^T = \hat{r} = \sqrt{\Delta z}(S_1 S_2)^{-1}\discq$ we have that the Hamiltonian in Eq.~\eqref{eq:app:H} takes the form
\begin{equation}
 \hat H_N = \tfrac12 \sum_{j = 1}^{N_0} \bigl(\discd_j\bigr)^{2} + \tfrac12\sum_{j=N_0+1}^N \omega_j (\bigl(\discd_j\bigr)^{2}+\bigl(\discp_j\bigr)^{2}),
\end{equation}
such that $\discp_j \approx \p_j$ and $\discd_j\approx \d_j$ as $\hat{H}\approx \hat{H}_N$. We will therefore not distinguish between $\discp_j$ and $\p_j$ and $\discd_j$ and $\d_j$ outside of this section.

\subsection{Discrete approximations}
With this, we can read off the discrete approximation to the wave functions $\fjp$ and $\fjd$ relating $\discp$ and $\discpp$ or correspondingly $\discd$  $\discdd$ as the rows of $S =  S_1 S_2 $ which is of block structure, i.e.,$S = S^\phi\oplus S^\rho$. Specifically we find
\begin{align}
\fjp(z_k) \approx \sqrt{\Delta z}^{-1}S^\phi_{ j+N_0, k}
\end{align}
and 
\begin{align}
\fjd(z_k) \approx \sqrt{\Delta z}^{-1} S^\rho_{ j+N_0, k }\ .
\end{align}

Note that when relating $\discp$ and $\discpp$ or $\discd$ and $\discdd$ we included a factor $\sqrt{\Delta z}$. The inclusion of this factor allows to change the commutation relations from $[\discdd_j,\discpp_k] = i\delta_{j,k}/\Delta z \rightarrow i\delta(z_j-z_k)$ to $[\discd_j,\discp_k] = i\delta_{j,k} \rightarrow i\delta_{j,k}$ as one would expect from the discrete canonical eigenmodes of the system.
Let us furthermore observe that the relation to the real-space correlators are given by
\begin{align}
 \langle \pp(z_j) \pp(z_k) \rangle \approx \langle \discpp_j \discpp_k \rangle = \Delta z^{-1} \sum_{j',k'=1}^N S^{-1}_{j,j'}S^{-1}_{j,j'} \langle \discp_{j'} \discp_{k'} \rangle,
\end{align}
where we exploited that $S$ has a block-diagonal structure and the inverse scaling in the discretization step $\Delta z$ should be noted.

\begin{figure}
\includegraphics[trim = 3cm 8cm 3cm 8cm, clip,width=0.45\textwidth]{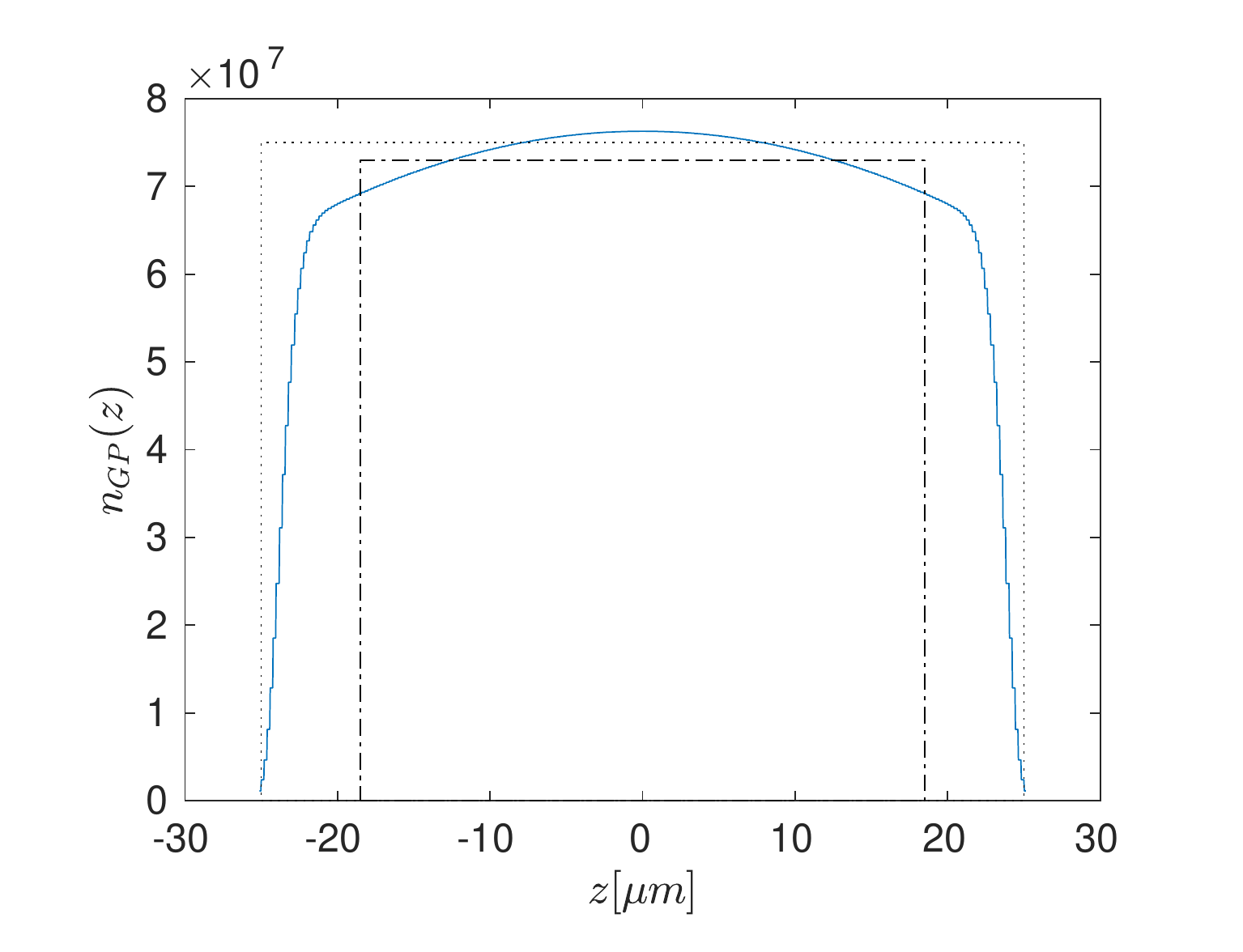}
\includegraphics[ trim = .2cm 0cm 3cm 15.5cm, clip, width=0.45\linewidth ]{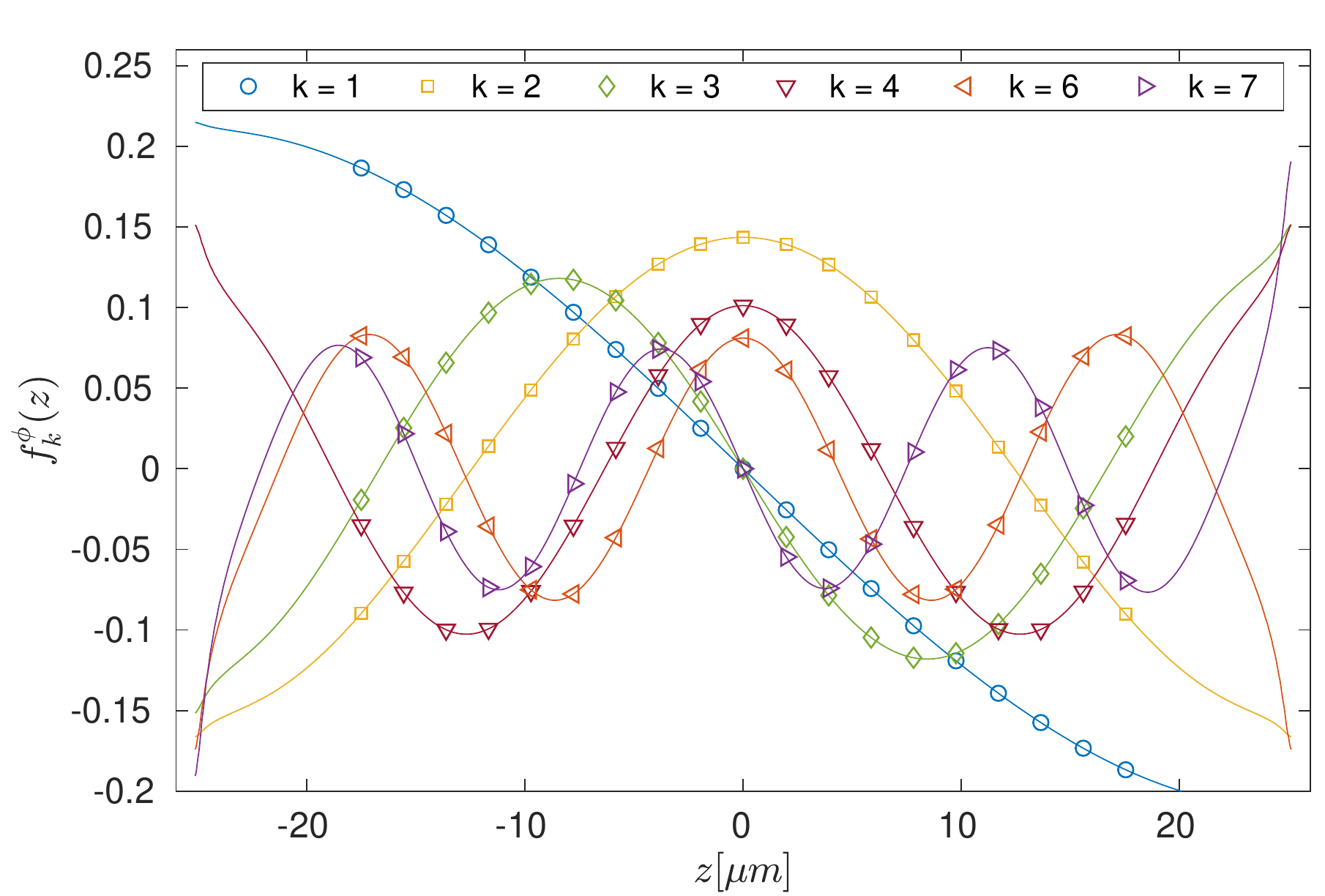}
\caption{
On the left we show the Gross-Pitaevskii profile $\nGP$ reflecting the setting of the experiment where the box trapping potential is finite and is superposed with an additional weak harmonic potential both of these features lead to $\nGP$ not being perfectly homogeneous. The dotted rectangle indicates a region of width \SI{50}{\micro\meter} such that at the edges the profile amounts to $5\%$ of the peak density.
The dashed-dotted region corresponds to the window where data is typically taken -- here the profile is relatively homogeneous.
On the right, we plot low-lying eigenmode functions taking the values $f^\phi_j(z_k) \approx \sqrt{\Delta z}^{-1} S^\phi_{ j, k}$ which show oscillatory behavior similar to the analytical solution to the Luttinger liquid model that can be obtained for the homogeneous profile.}
\label{fig:app:GP}
\end{figure}

\section{Detailed formulation of the recovery procedure}
\label{sec:app:recoverydetails} 

\begin{figure}
\includegraphics[trim = 3cm 8.cm 2cm 8.7cm, clip, width =.45\linewidth]{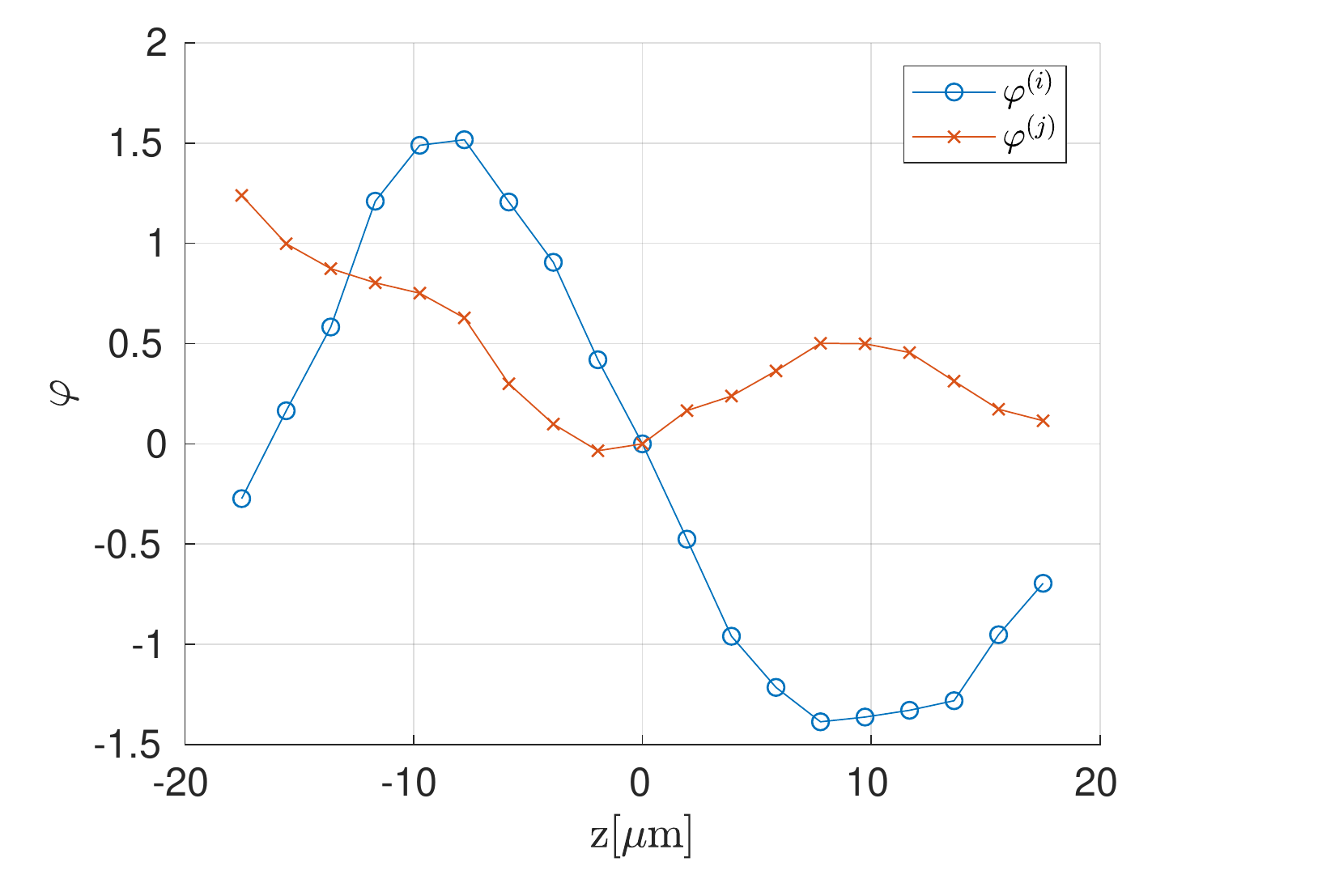}
\includegraphics[trim = 3cm 8.cm 2cm 8cm, clip, width=0.5\linewidth]{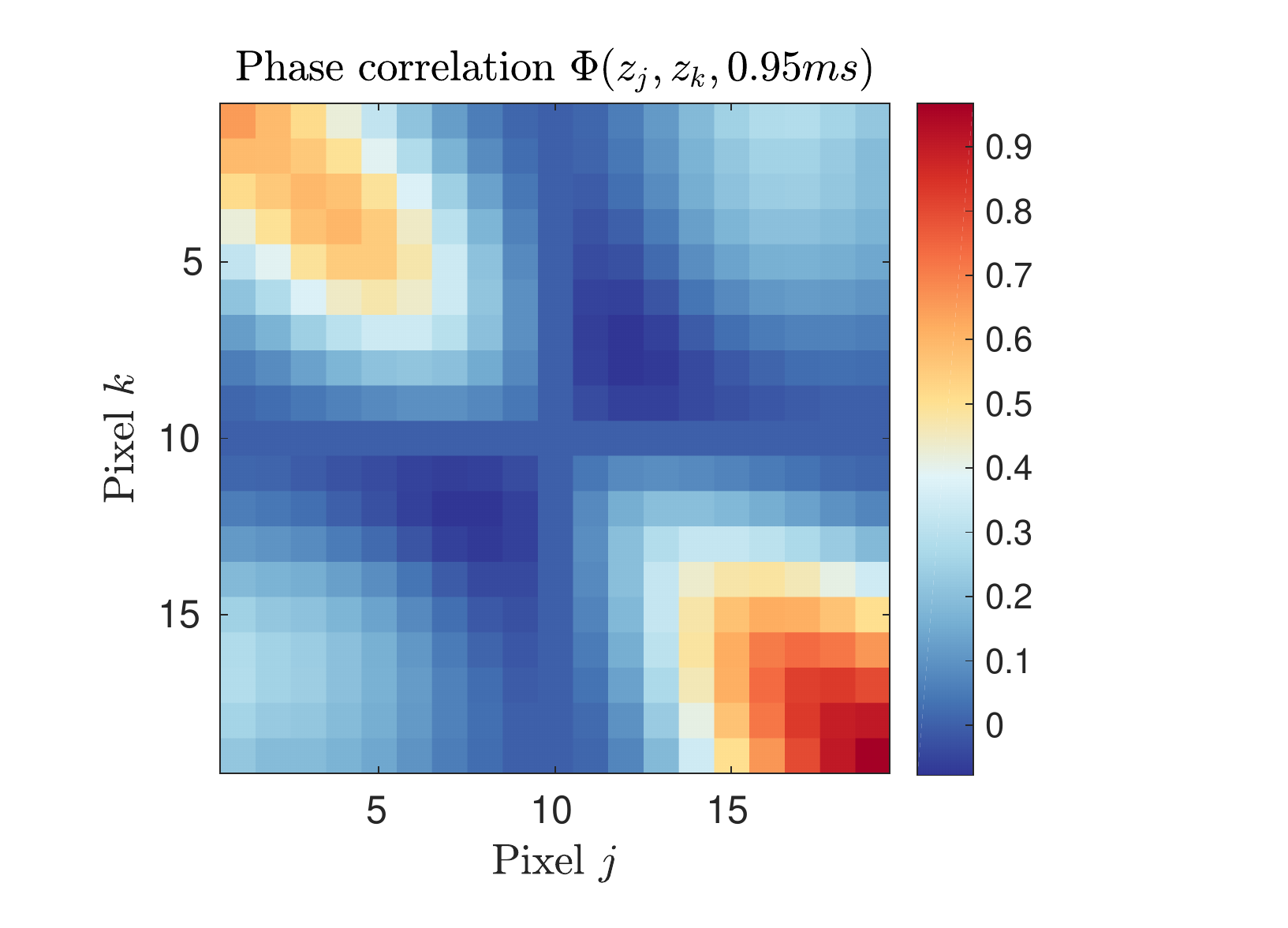}
\caption{On the left we plot an example of two arbitrarily chosen referenced phase profiles $\varphi^{(i)}$ and  $\varphi^{(j)}$ obtained in the experiment referenced to the center of the considered portion of the cloud. On the right we show an example of the measured values of $\tilde\Phi(z_a,z_b,t)$ extracted from $n_{\rm Sample}\approx200$ profiles. The referenced second moments vanish trivially if $z_a$ or $z_b$ equals the reference point and increase with increasing distance to $z_0$.}
\label{fig:protocol}
  \end{figure}
In this section, we describe the data analysis and formulate the reconstruction procedure with additional details.
As described in Refs.~\cite{onsolving, Recurrence,van2018projective} through matter-wave interferometry phase profiles $\varphi$ of the superfluid can be measured.
After the two superfluids were coupled with tunneling strength $J\approx\SI{ 3.5}{\hertz}$ for sufficiently long times, the separation potential is increased rapidly in about $\SI{1}{\milli\second}$.
The end of the ramp defines the initial time $t_0=\SI{0}{\milli\second}$ for the quench evolution that follows.
The gas can then be held for a specific time $t$ which defines the time during which the system evolves under the quench Hamiltonian in Eq.~(1).
Here we focus on the referenced second moments obtained from the profiles, for  which the corresponding physical observable is
\begin{align}
  \Phi(z,z',t)&=\expval{[\pp(z,t)-\pp(z_0,t)][\pp(z',t)-\pp(z_0,t)]}\ ,
  \label{eq:app_phi}
  \end{align}
where $z_0$ denotes a fixed reference point in the system.

For each hold time $t$ about  $n_{\rm Sample}\approx200$ phase profiles were obtained, this number is limited by the stability of the setup, but can be increased if fewer hold times are considered in total.
Fig.~\ref{fig:protocol} shows an example of two of many phase profiles used in the analysis.
In the data analysis, data from $\NumP = 19$ central pixels is used which corresponds to about $60\%$ of the cloud as each pixel has size $\ell = \SI{1.95}{\micro\meter}$, so the total size observed is about $\SI{37}{\micro\meter}$.
The positions of the pixels are
\begin{align}
  z_a  =  ( a - 10 )\ell
\end{align}
and the reference point is chosen as $z_0 =\SI{0}{\micro\meter}$, i.e., in the middle of the cloud.
We consider referenced second moments here, as by this we are able to consistently remove any offset phase between two different measured profiles. In practice we subtract at each pixel $a$ the central phase profile value $\varphi^{(i)}(z_0)$.
The experimental estimate for Eq.~\eqref{eq:app_phi} is then
\begin{align}
   \Phi_{\rm est}(z_a,z_b,t)&=\frac{1}{n_{\rm Sample}}\sum_{i=1}^{n_{\rm Sample}} \bigl(\varphi^{(i)}(z_a) - \varphi^{(i)}(z_0)\bigr)\bigl(\varphi^{(i)}(z_b) - \varphi^{(i)}(z_0)\bigl)\ .
  \label{eq:estimate_varphi}
  \end{align}
In terms of the eigenmodes this reads
\begin{align}
 \Phi( z_a,z_b,t )   &=\frac{1}{2}\sum_{j,k=1}^\N(   f^\phi_j(z_a)-   f^\phi_j(z_0))(   f^\phi_k(z_b)-   f^\phi_k(z_0))\ \expval{\{\p_{j}(t),\p_{k}(t)\}}:=  \sum_{j,k=1}^N  \fjk  \Vpp_{j,k}(t)\ .
  \end{align}
Using the expressions for the time evolution, we get
\begin{align}
\label{eq:data_to_quadratures2}
\begin{split}
\Phi( z_a,z_b,t )  =&\sum_{j,k=1}^\N \fjk \bigl(\cos(E_{j} t ) \cos(E_{k} t )\Vpp_{j,k}  +\sin(E_{j} t ) \sin(E_{k} t )\Vrr_{j,k}\bigr)\\
 & -\sum_{j,k=1}^\N (\fjk+\fkj)\cos(E_{j} t ) \sin(E_{k} t )\Vpr_{j,k}\ .
 \end{split}
\end{align}

It must be noted that the measured value at a pixel $z_a$ does not exactly reflect the value of the field $\pp(z_a)$ but rather a \emph{convolution} of the field with a Gaussian function, i.e., it probes the value averaged over a patch of specific characteristic length.
More precisely, the experiment allows us only access to measurements of 
\begin{align}
  \tilde \pp(z_a) = \frac{\int_{-\RD}^{\RD} \di z^\prime\ e^{-\frac{(z^\prime-z_a)^2}{2\sigma^2}} \pp(z^\prime) }{ \int_{-\RD}^{\RD} \di z^\prime\ e^{-\frac{ (z^\prime-z_a)^2}{2\sigma^2}} }
\end{align}
and we define the correspondingly convoluted wave function 
\begin{align}
  \tilde f^\phi_j(z_a) =  \frac{\int_{-\RD}^{\RD} \di z^\prime e^{-\frac{(z^\prime-z_a)^2}{2\sigma^2}} f^\phi_j(z^\prime) }{ \int_{-\RD}^{\RD} \di z^\prime e^{-\frac{ (z^\prime-z_a)^2}{2\sigma^2}} }\ .
  \label{eq:SM_f_conv}
\end{align}
For the considered experimental setup we find the estimation $\sigma \approx \SI{3}{\micro\meter}$.
In order to include the convolution in the reconstruction it then suffices to use the convoluted wave functions and set
\begin{align}
\tilde \Phi( z_a,z_b,t )   &=\frac{1}{2}\sum_{j,k=1}^\N(  \tilde f^\phi_j(z_a)-  \tilde f^\phi_j(z_0))(  \tilde f^\phi_k(z_b)-  \tilde f^\phi_k(z_0))\ \expval{\{\p_{j}(t),\p_{k}(t)\}}:=  \sum_{j,k=1}^N  \tilde \fjk  \Vpp_{j,k}(t)\ .
\label{eq:data_to_quadratures_conv}
  \end{align}

\subsection{Recovery procedure}
In the implementation, we vectorize the covariance matrix $V$ such that each block is a vector, i.e., $ v^{\phi \phi} =\text{vec}(\Vpp)$ etc. and define $v=v^{\phi \phi}\oplus v^{\phi\rho}\oplus v^{\rho\rho} \in\mathbb R^{3\N^2}$ and use the notation $v={\rm vec}{(V)}$.
Formula \eqref{eq:data_to_quadratures2} shows that for each input $(z_a,z_b,t)$ we can find a vector $w\in\mathbb R^{3\N^2}$ such that $\Phi(z_a,z_b,t)=w^T v$.
For a fixed time $t_{i_k}$ we then collect the from the measured data extracted second moments $\Phi_{\rm est}(z_a,z_b,t_i)$ in a vector $b_k\in\mathbb{R}^{\NumP^2}$ and construct the corresponding vectors $w$ and collect them as rows in a matrix $A_k\in \mathbb{R}^{\NumP^2 \times 3\N^2}$. 
Doing this for all $n$ times $t_{i_1},\dots, t_{i_n}$ which are used for the reconstruction as input, we then stack all $b_k$ and $A_k$ into a large vector $b$ matrix $A$ correspondingly, i.e.,
\begin{align}
  A=\begin{bmatrix} A_1\\\vdots\\ A_{n}\end{bmatrix}\,,\quad b=\begin{bmatrix} b_1\\\vdots\\ b_n\end{bmatrix}\ .
\end{align}
We furthermore define at each time step a diagonal matrix $W\in\mathbb{R}^{\NumP^2\times \NumP^2}$ which contains the inverse statistical errors of the experimental measurement of the second moments $W^{(k)}_{(a,b),(a,b)} = 1 / \sigma_{\rm est }(\Phi(z_a,z_b,t_k))$ and collect all $W^{(k)}$ in on large block-diagonal matrix $W$ in order to define a more uniform target function for the optimization.
With this definition we aim at minimizing the vector Hilbert Schmidt-norm 
\begin{align}
\Theta = \|WA \,{\rm vec}{(V)} -Wb\|_2
\label{eq:SM_ls}
\end{align}
subject to the semi-definite constraint
\begin{align}
V+\tfrac12 i\Omega\succeq 0\ ,
\end{align}
where in the main text we have introduced the notation $\mathcal A(V) = A\,{\rm vec}{(V)}$.
The numerical reconstruction has been implemented with use of the \texttt{cvx} package.
The standard theory of semi-definite programming shows that there is always a unique solution $v_{\rm Opt}$ to this optimization problem.
Unfolding the vectorization yields the reconstructed covariance matrix $V_{\rm Opt}$.

As a final remark, it is interesting to note that positivity constraints (imposing that the density operator is positive semi-definite) 
similar to the Heisenberg constraint (reflecting the Heisenberg uncertainty principle as a semi-definite constraint) 
characterizing bosonic covariance matrices can significantly increase stability of least squares reconstructions \cite{Kalev2015}.
In fact, wide classes of recoveries with a positivity constraint \cite{Kalev2015} can be interpreted as compressed sensing schemes 
\cite{Compressed}. In this context, is important to stress that the semi-definite 
constraint $V+\tfrac12 i\Omega\succeq 0$ readily implies that $V>0$, so that $V$ is strictly positive, so that the constraint of Ref.\
\cite{Kalev2015} is readily enforced. Hence, it is interesting to see that much of the intuition on the positive cone for 
density operators carries over to the Heisenberg cone for covariance matrices. Further explorations of seeing our scheme as
a compressed sensing scheme will be left to future work.

\section{Simulation of the reconstruction procedure }
\label{sec:app:simulateddata}
Various aspects of the reconstruction can be modeled by considering a thermal state of the effective Hamiltonian of the strongly coupled condensates as discussed in the main text.
Thermal correlations in the discretized model can be obtained either by considering the exact formulas from 
Refs.~\cite{audenaert2002entanglement, banchi2015quantum} or by classical phase approximation \cite{zinn1996quantum}.
In the following we consider the latter and study the effects of finite sample size and finite measurement resolution.
We denote the real-space phase fluctuation functions by 
\begin{equation}
	(\Gamma^{\phi\phi})_{a,b} = \langle \pp(z_a)\pp(z_b) \rangle 
\end{equation}
and in classical field approximation we can calculate these via $\Gamma^{\phi\phi} \approx (H_\phi(J\neq 0))^{-1} / k_B T $ \cite{zinn1996quantum}. 
Together with the correlations for density fluctuations, we can propagate these under the quench Hamiltonian.
Fig.~\ref{fig:Phi_s} shows the correlations $\Gamma^{\phi\phi}$ and additionally the effect of the referencing which removes the running phase, of the convolution which comes from the measurement resolution and finally discretization due to a finite amount of pixels.

\begin{figure}[h]
\includegraphics[trim = 4cm 8.5cm 4cm 8cm, clip, width=0.24\linewidth]{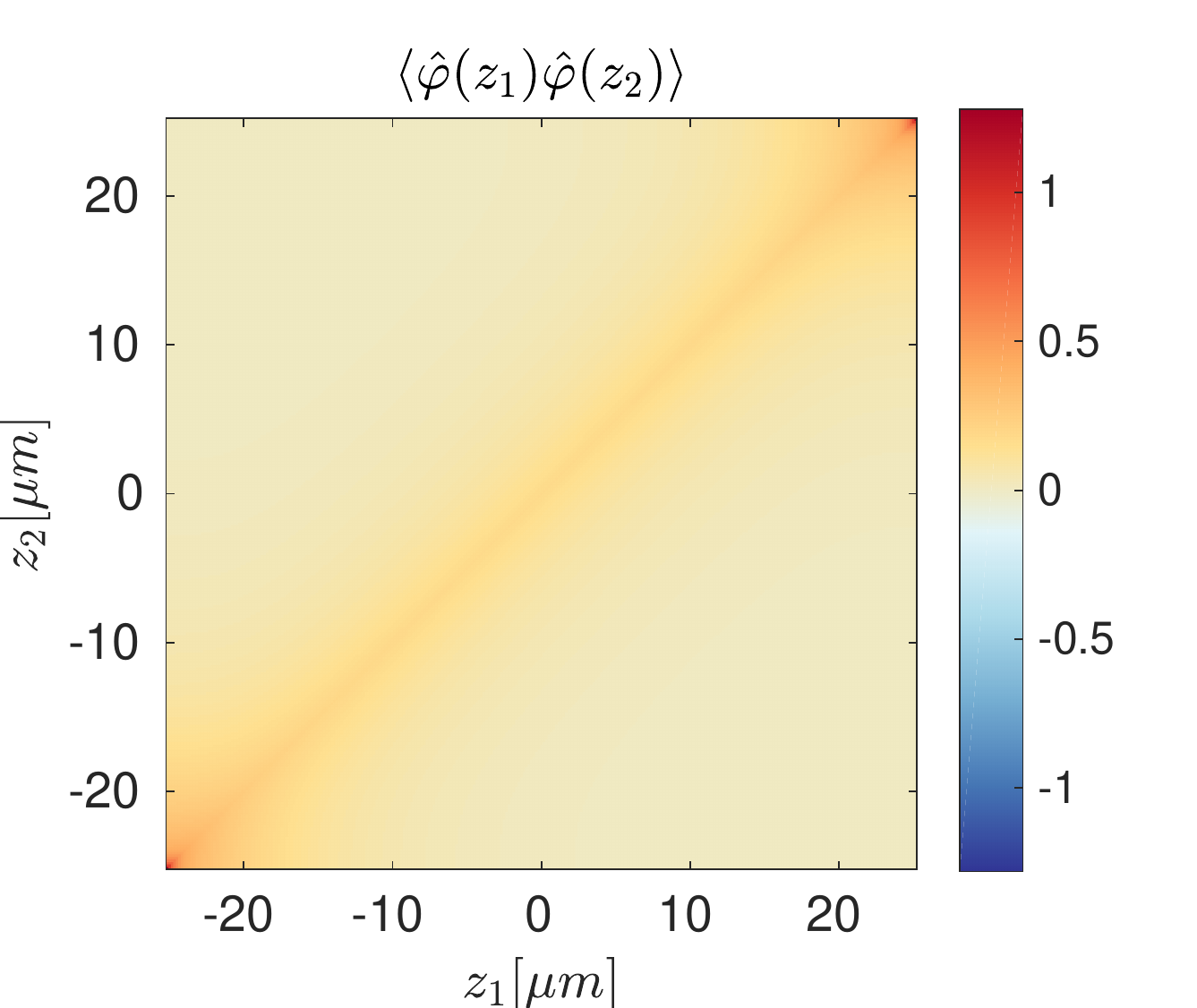}
\includegraphics[trim = 4cm 8.5cm 4cm 8cm, clip, width=0.24\linewidth]{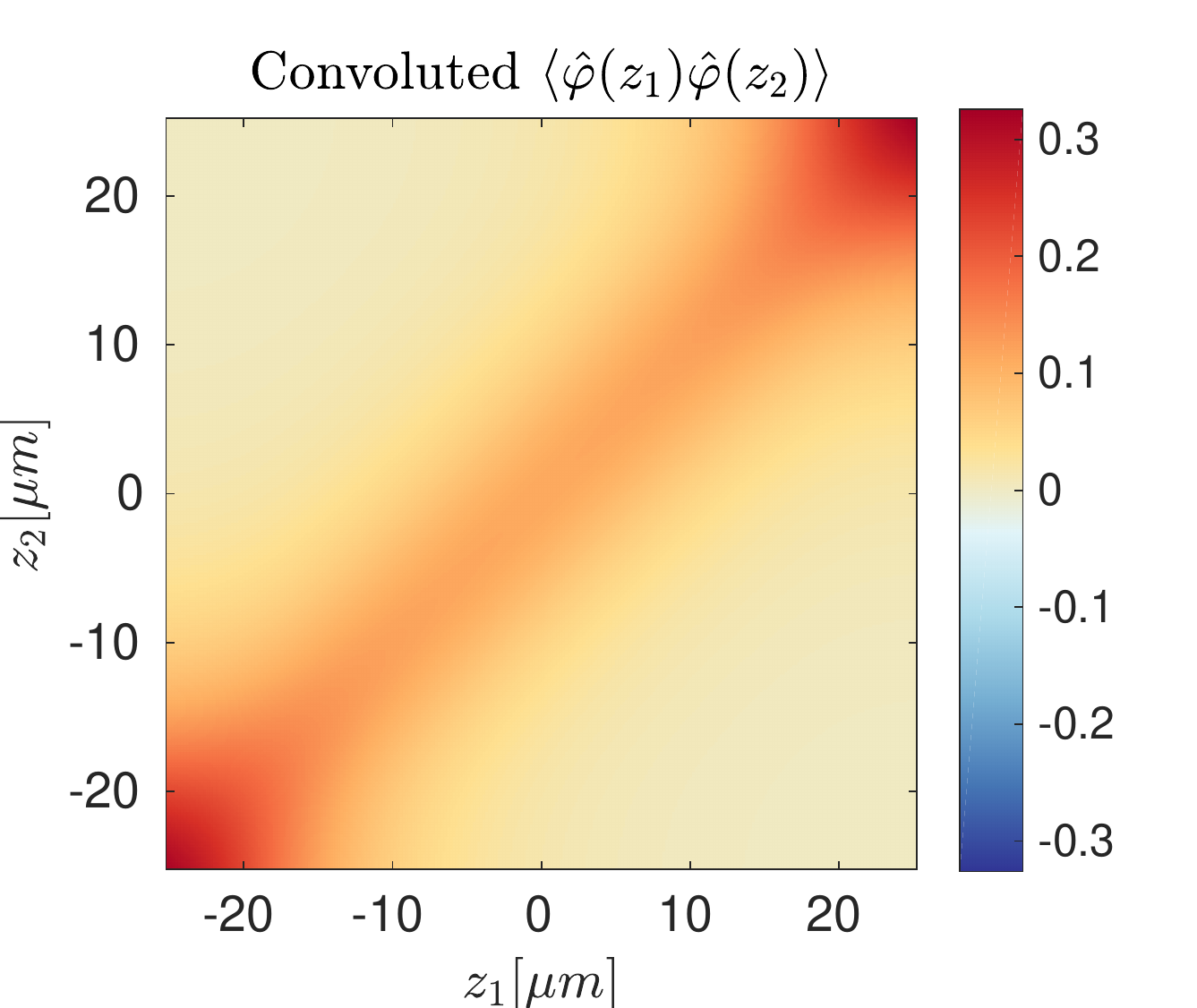}
\includegraphics[trim = 4cm 8.5cm 4cm 8cm, clip, width=0.24\linewidth]{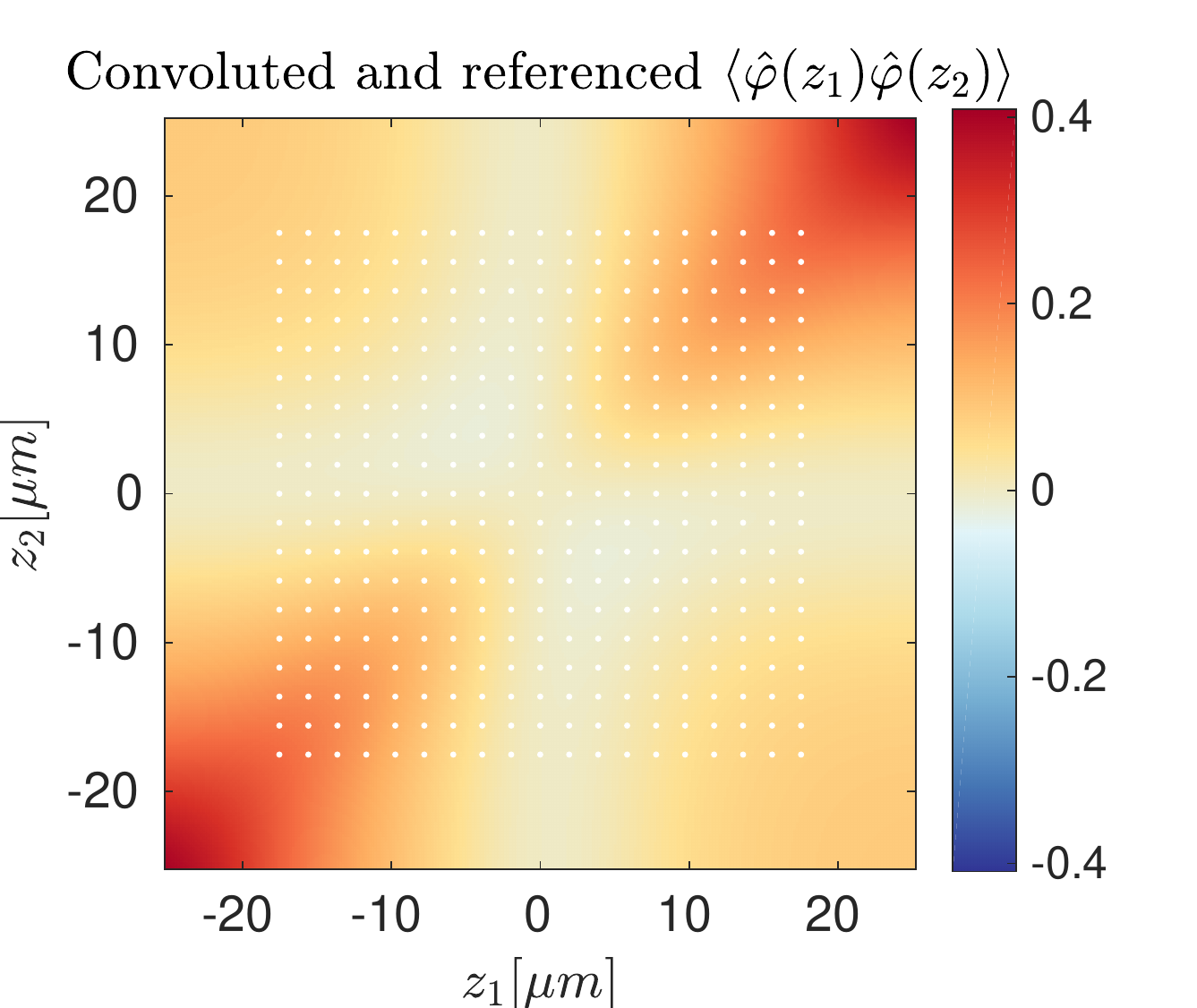}
\includegraphics[trim = 4cm 8.5cm 4cm 8cm, clip, width=0.24\linewidth]{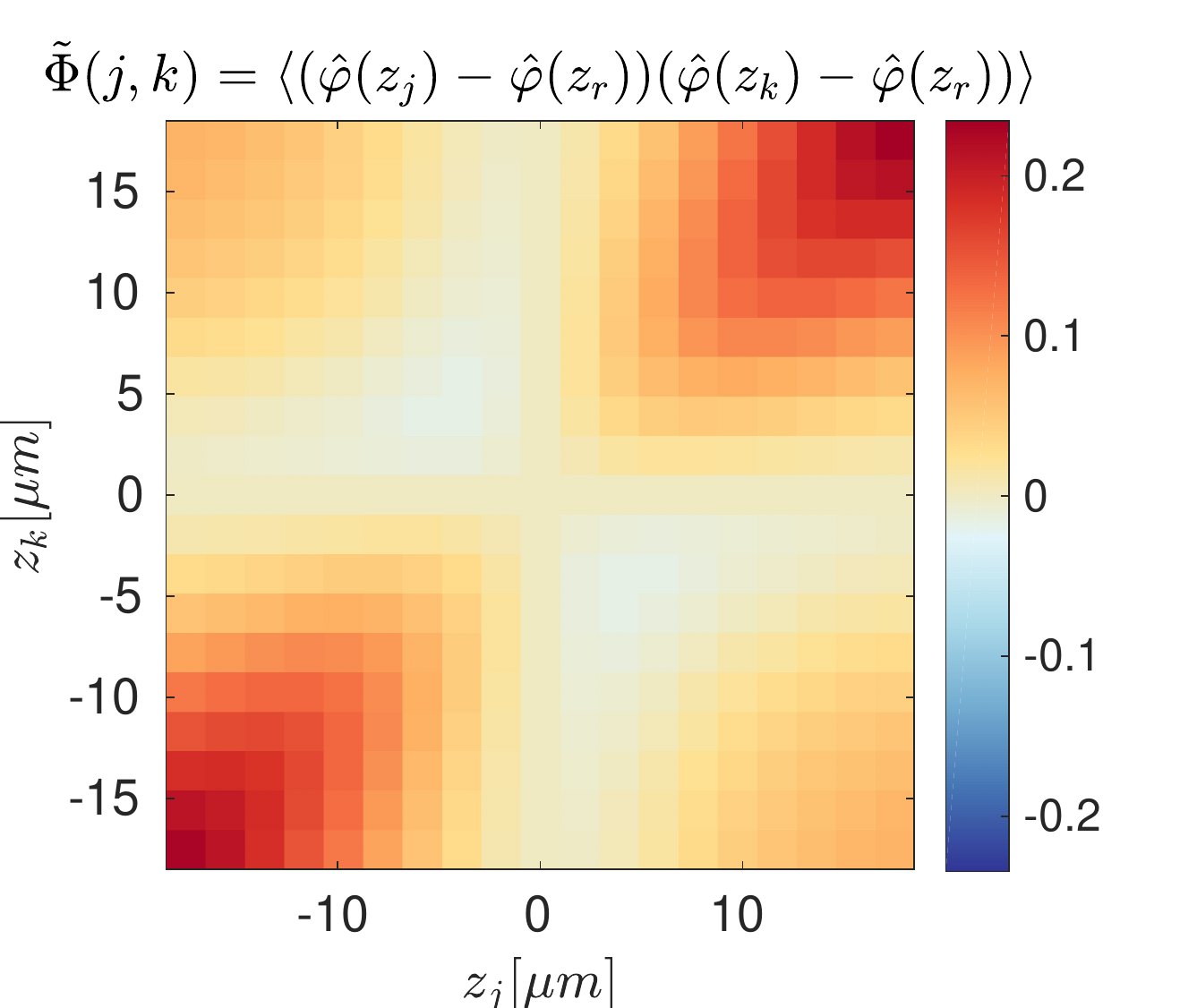}
\caption{ Phase correlations in real-space. From the left: direct values, convoluted correlations, convoluted and referenced correlations (pixel positions indicated by white dots) and finally the convoluted referenced observable evaluated at pixel positions. }
\label{fig:Phi_s}
\end{figure}

\subsection{Influence of finite sample size}
At each time the phase correlations can be used as a positive matrix parametrizing a (classical) Gaussian distribution from which single-shot random profiles can be sampled.
After convolution these can be seen to correspond to the direct observable of the interferometry which is useful to assess the systematic imperfections of our procedure.
Here we study two possible aspects.
By resampling with $n_{\rm Sample}=200$ and $n_{\rm Sample}=2000$ we study the sensitivity of our reconstructions to statistical fluctuations in the experiment.
Secondly, we study the real-space correlations and the convolution of these to see how finite measurement resolution impacts the information that can be obtained from our procedure.

We have simulated the $\hat H_N$ for $N=400$ and $J/\hbar = 2\pi\times\SI{3.5}{\hertz}$ obtaining the correlation functions of the thermal state at $T=\SI{40}{\nano\kelvin}$.
Using the phase-phase correlation functions at different times we have resampled the profiles.
After referencing the profiles the resampled phase-phase correlation functions were obtained.
We then perform a recovery at the times indicated in Fig.~\ref{fig:resampling1} which shows that finite sample size of about 200 experimental runs for each time constrains the possibility of recovering the phase locking in its full extent.

\begin{figure}
\includegraphics[trim = 3.5cm 8.1cm 4cm 8cm, clip, width=0.45\linewidth]{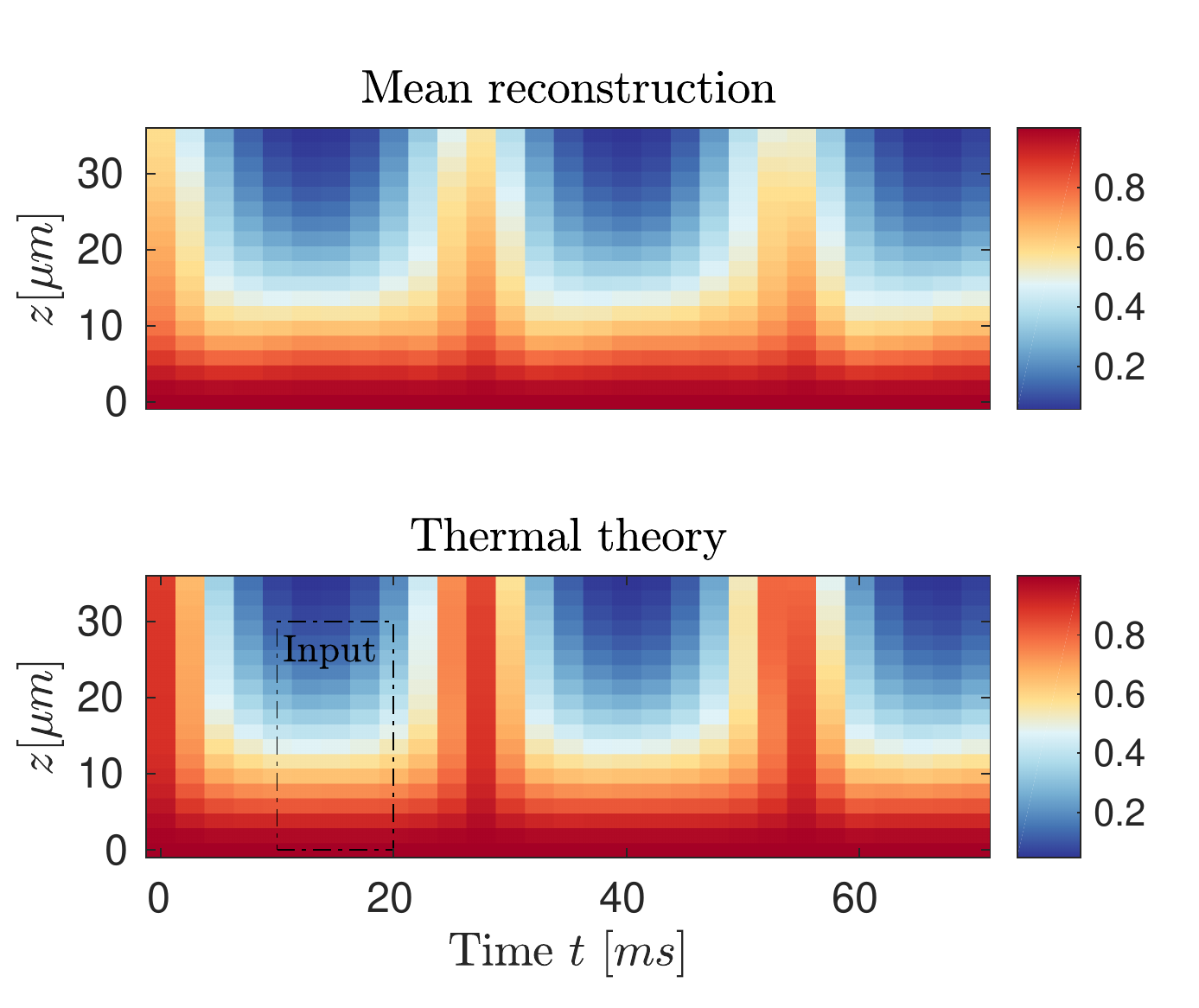}
\includegraphics[trim = 3.5cm 8.1cm 4cm 8cm, clip, width=0.45\linewidth]{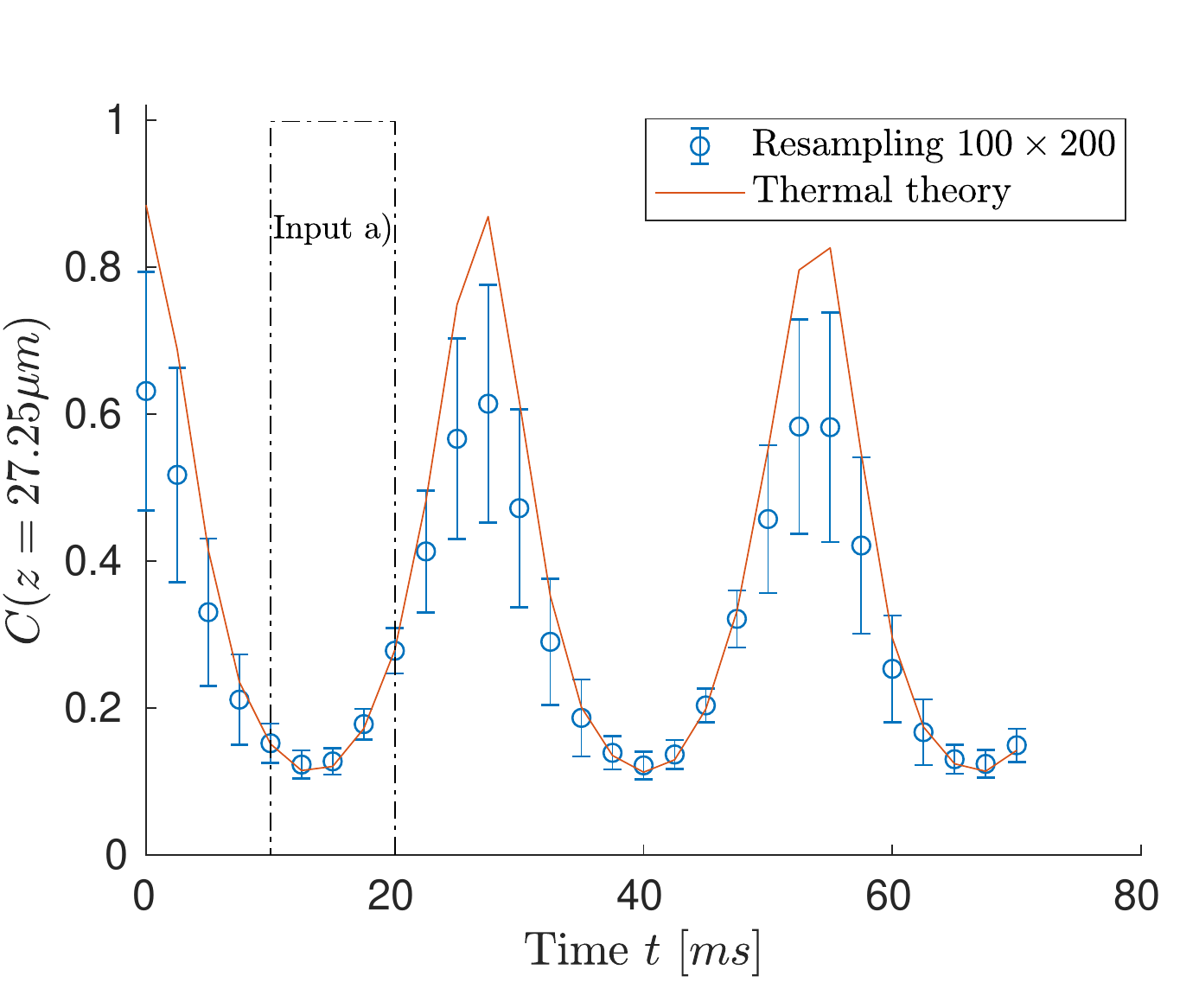}
\caption{We present the results of $100$ reconstructions obtained based on 200  phase profiles sampled from the thermal state  of the state preparation Hamiltonian with a phase coupling $J/\hbar =2\pi\times \SI{3.5}{\hertz}$ and $T =\SI{40}{\nano\kelvin}$.
On the left we show the averaged correlator $C$ as was done in the main text (the input window corresponds to the input region a) in Fig.~3 in the main text).
We find that the height of the reconstructed revivals based on estimators with sample size $n_{\rm Sample}=200$ is significantly lower than for the exact correlations obtained as an ensemble average from the thermal state.
On the right we present a cut at $z = \SI{ 27.25}{\micro\meter}$ for the thermal state (in red) and the reconstructions from the resampled data (blue points) where the error-bars indicate the standard deviation of the $100$ reconstructions.
This uncertainty analysis shows than based on a single sample of $n_{\rm Sample}=200$ profiles available experimentally the heights of the reconstructed revivals may fluctuate strongly.
The large standard deviation for points at the revival indicate that  both large and small revival heights can be obtained depending on random fluctuations of the sampled phase profiles. 
}
\label{fig:resampling1}
\end{figure}
\begin{figure}
\includegraphics[trim = 3.5cm 8.4cm 4cm 8cm, clip, width=0.45\linewidth]{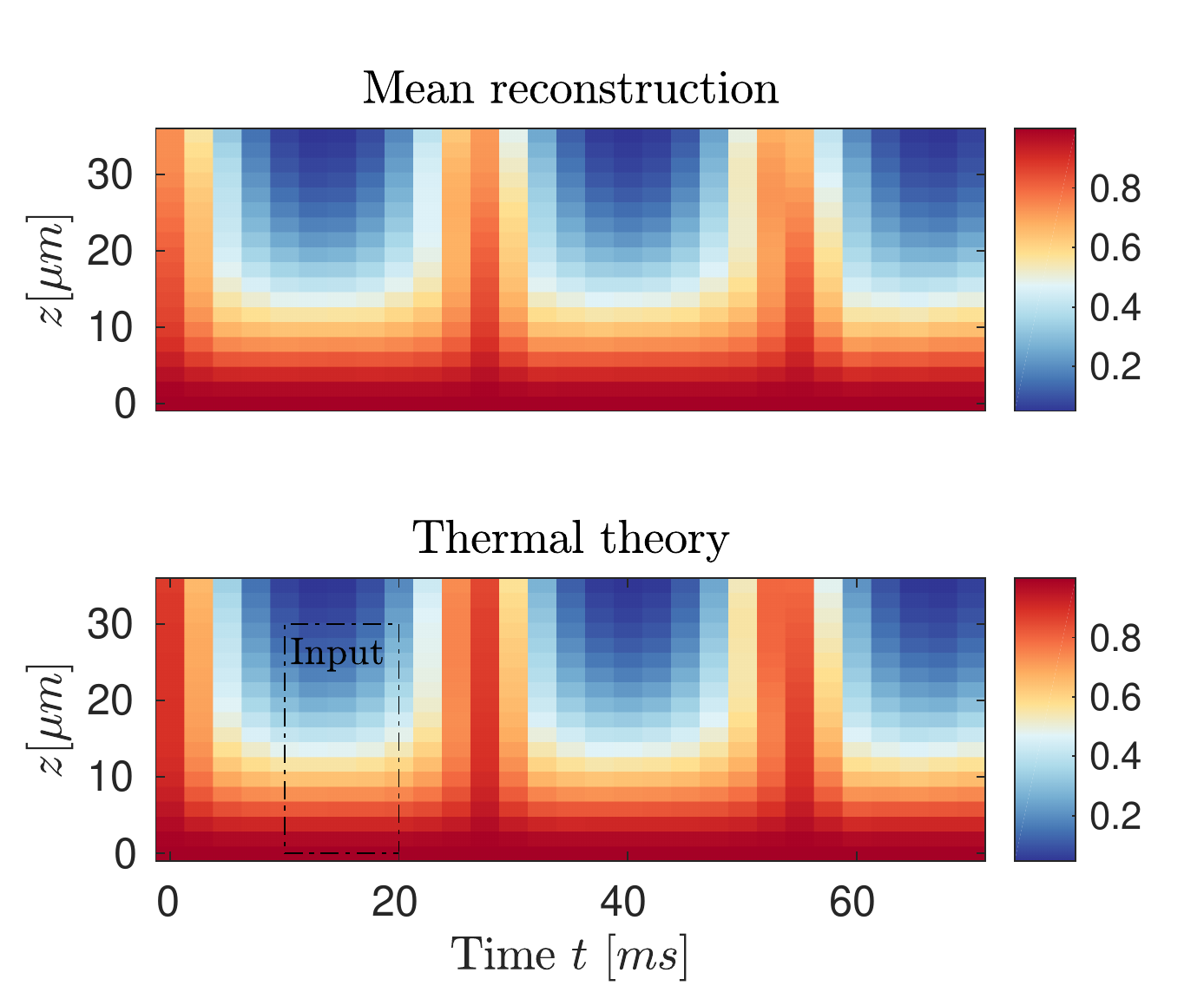}
\includegraphics[trim = 3.5cm 8.4cm 4cm 8cm, clip, width=0.45\linewidth]{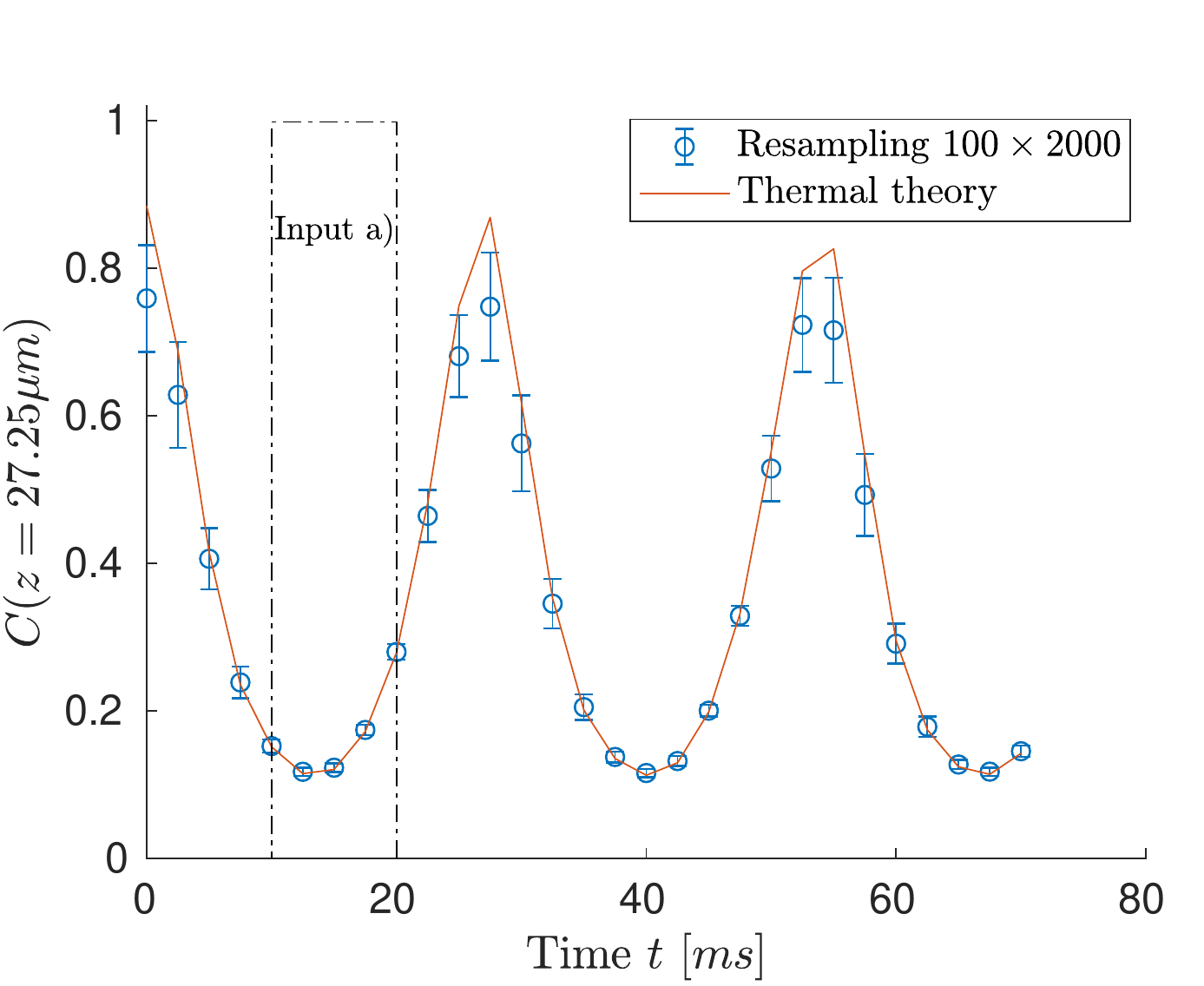}
\caption{As above, but using $n_{\rm Sample}=2000$ for which the estimators of the two-point correlation functions based on the sampled phase profiles should be close to the true values. We find that our method can very reliably reconstruct the state, including revival heights, and hence identify the finite sample size in the experiment $n_{\rm Sample}=200$ as the main source of inaccuracy of our reconstructions.}
\label{fig:resampling2}
\end{figure}

\subsection{Influence of finite measurement resolution}
The measured phase fluctuations at a given pixel are in fact a convolution which spreads into the neighboring pixels too.
Its primary effect is introducing a frequency cutoff, as the higher modes oscillate quickly and are averaged out.
The secondary effect is introducing non-universal additional artifacts into the correlations that come from the convolution and are a feature of the measurement setup.
We consider the thermal real-space covariance matrix.
It is diagonal in the eigenmodes of the initial strongly coupled Hamiltonian, see Fig.~\ref{fig:smearing} left.
The convolution introduces however new correlations that are not present in the state and are an artifact of the coarse-graining, Fig.~\ref{fig:smearing} second from the left.
If we consider the post-quench Hamiltonian, the modes change slightly due to the non-homegenous GP profile and the covariance matrix is slightly off-diagonal in these modes Fig.~\ref{fig:smearing} second from the right.
After the convolution again a cut-off is introduced, but also additional stray artifacts, Fig.~\ref{fig:smearing} right.

Thus, the convolution will introduce in an uncontrolled way additional artifacts at different times and hence the measured real-space second moments of phases will have a discrepancy incorporated by the finite measurement resolution.
This explains why in the data analysis, and the resampling simulation above, a perfect reconstruction is impossible.
We conclude that the sample size is a smaller limitation than the finite experimental resolution.
\begin{figure}
\includegraphics[trim = 5cm 8.5cm 5cm 8cm, clip, width=0.24\linewidth]{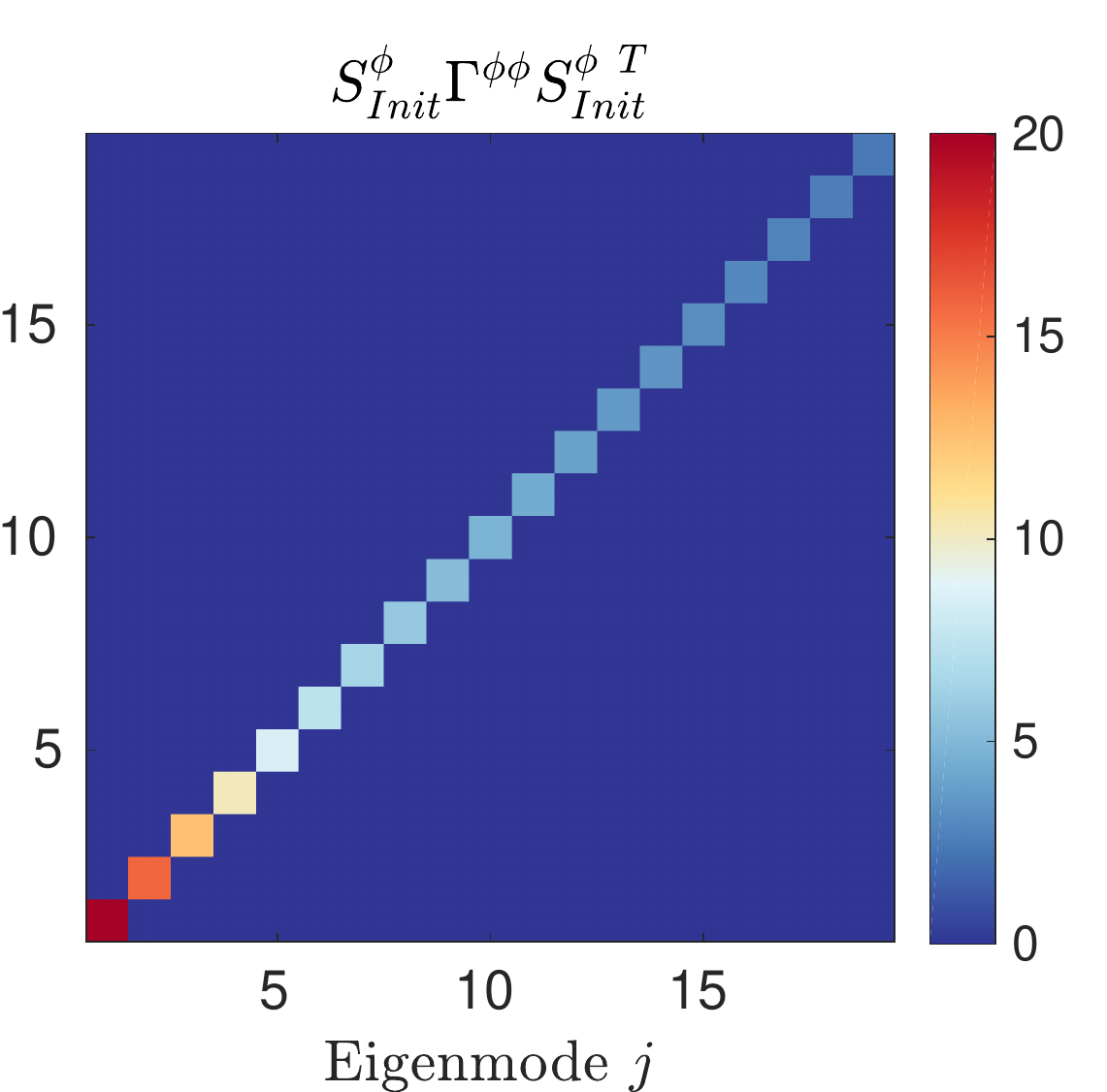}
\includegraphics[trim = 5cm 8.5cm 5cm 8cm, clip, width=0.24\linewidth]{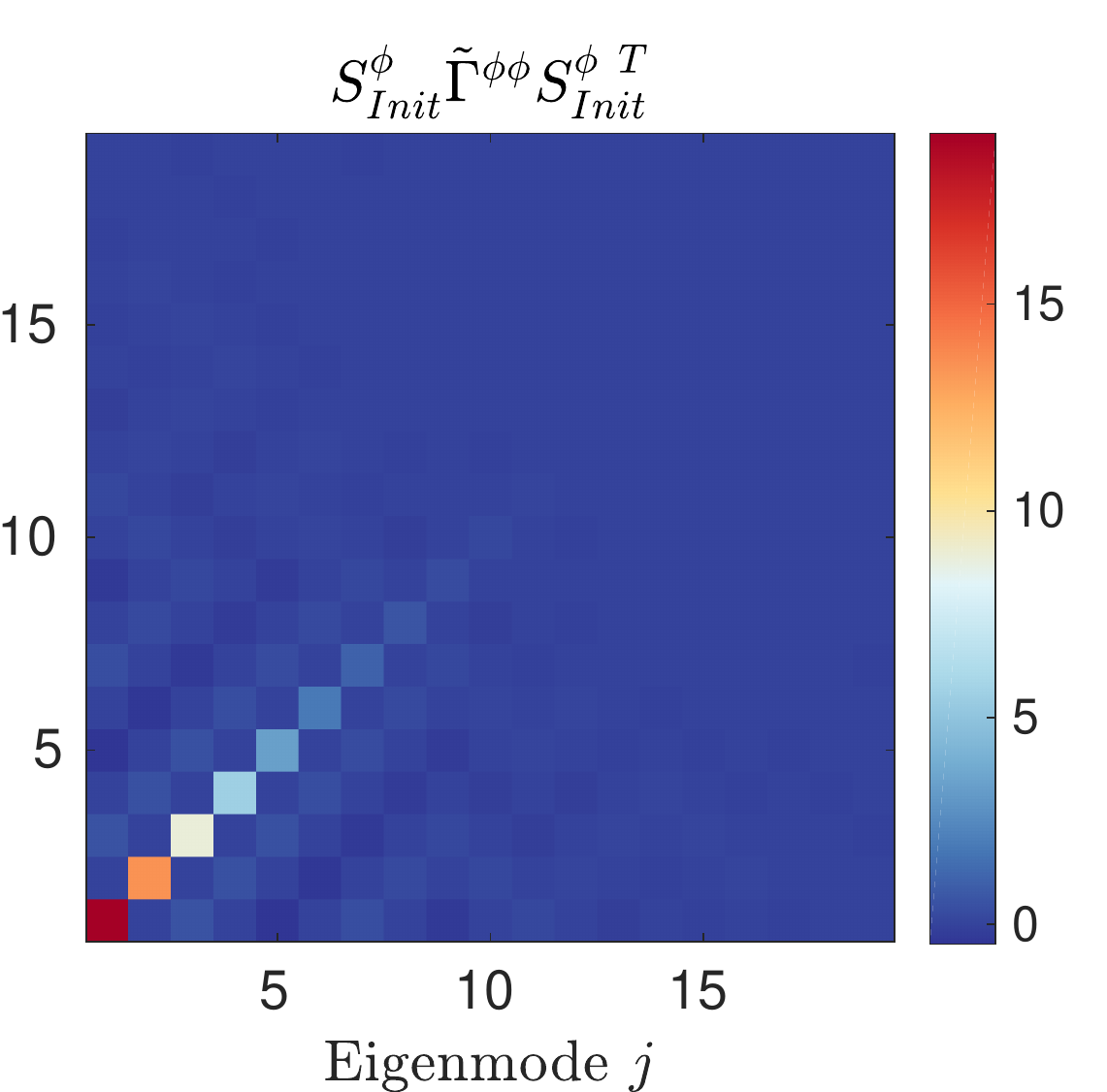}
\includegraphics[trim = 5cm 8.5cm 5cm 8cm, clip, width=0.24\linewidth]{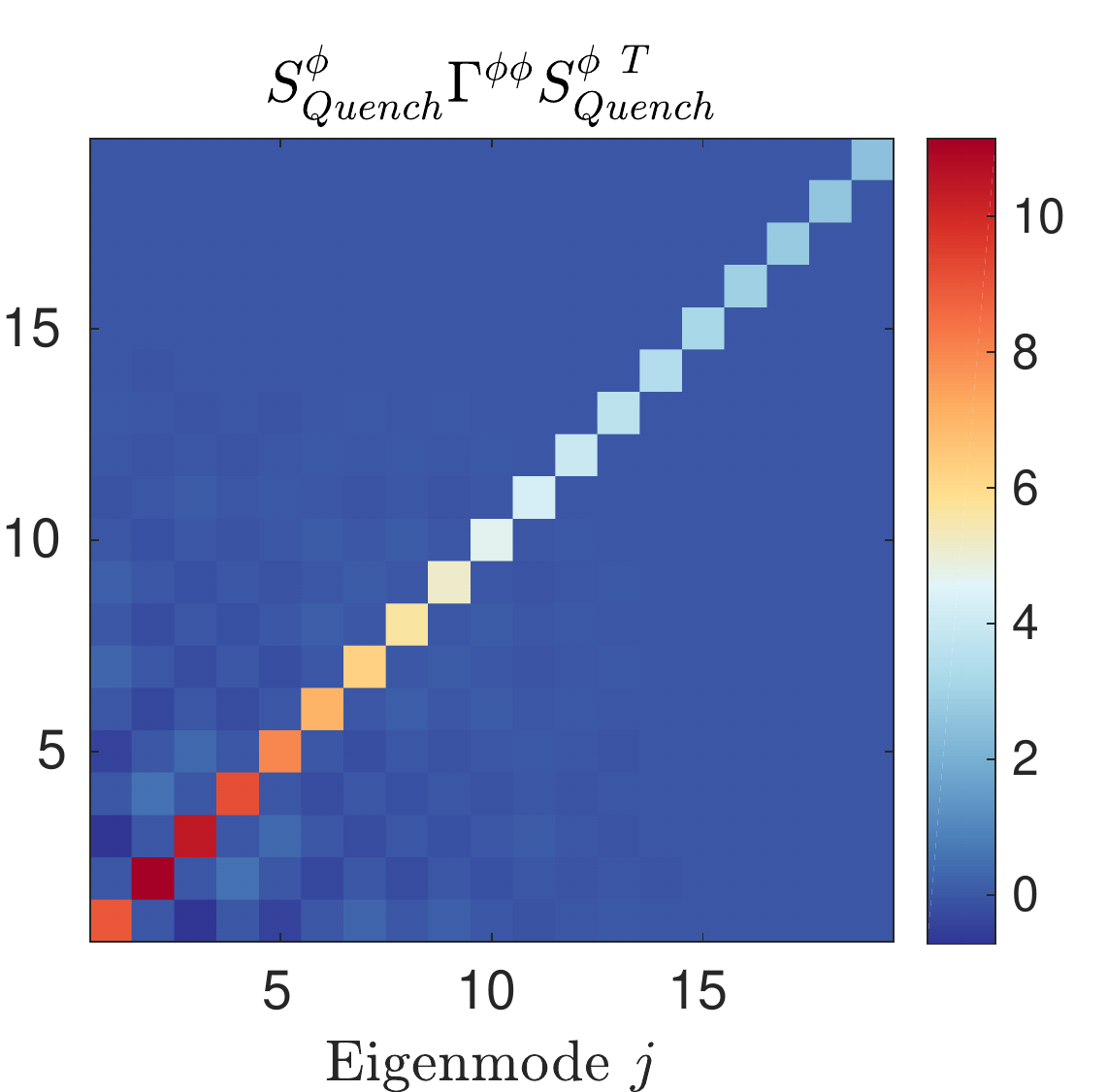}
\includegraphics[trim = 5cm 8.5cm 5cm 8cm, clip, width=0.24\linewidth]{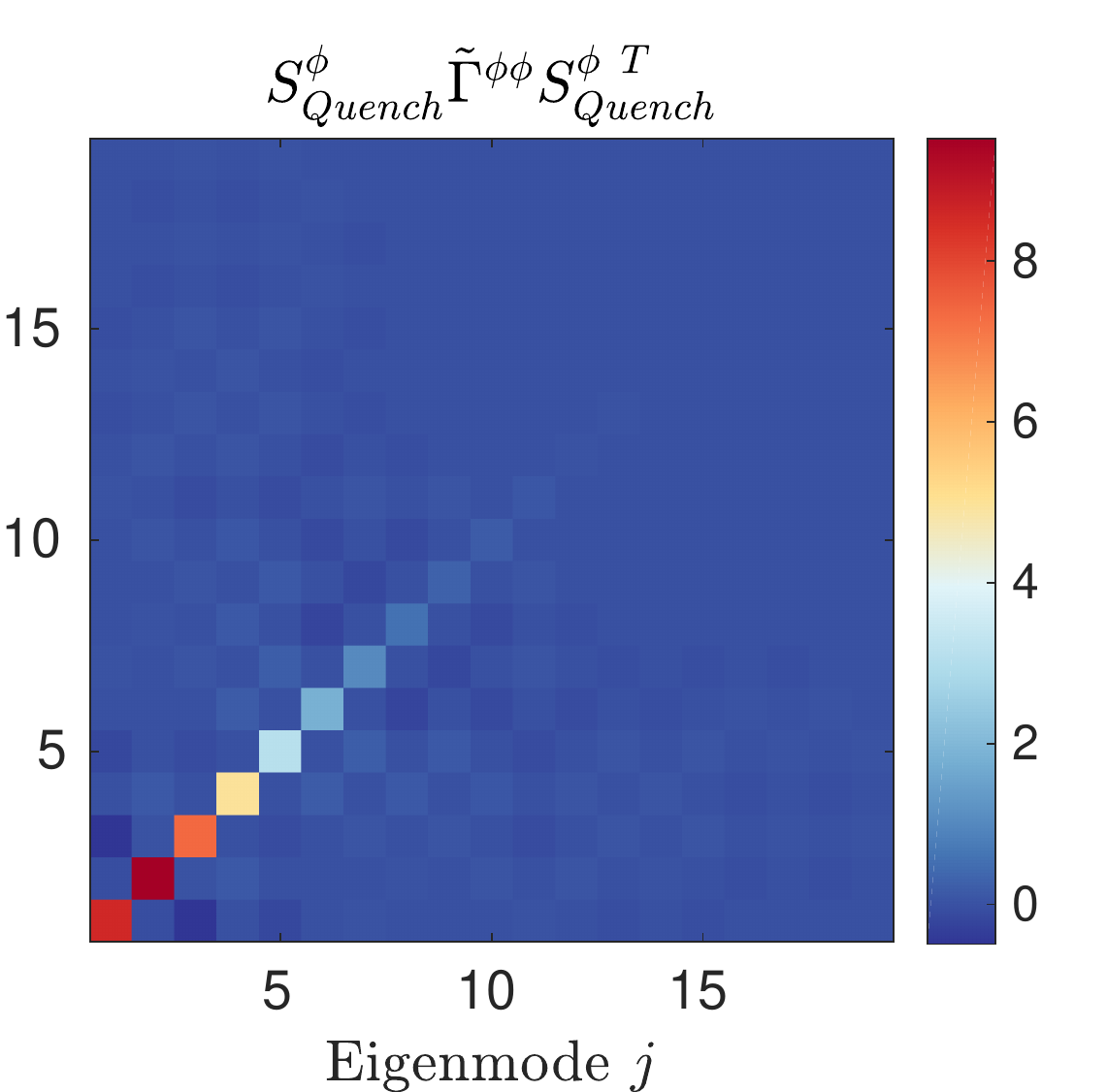}

\caption{ Real space correlations after rotating into the eigenmode space.
From left to right, the first plot shows the rotation of the real space covariance matrix $\Gamma^{\phi\phi}_{j,k}=\langle \pp(z_j)\pp(z_k)\rangle$ defined at discretization pixels rotated into the eigenmode space $S^\phi_{\rm Init}$ of the initial  Hamiltonian.
On all plots we show the lowest-lying eigenmodes and do not show the zero-energy mode considering only the eigenmodes relevant in the experiment.
Second  plot presents the Gaussian convolution of $\Gamma^{\phi\phi}$ denoted by $\tilde \Gamma^{\phi\phi}$ which takes into account the finite measurement resolution in the eigenmode space of the initial Hamiltonian.
Observe, that in both cases we obtain diagonal matrices, but the convolution introduces a cut-off for resolving the occupation of the higher modes.
The last two plots show the same comparison of $\Gamma^{\phi\phi}$ and $\tilde \Gamma^{\phi\phi}$ but now rotated with the eigenmodes $S^\phi_{\rm Quench}$ of the quench Hamiltonian.
Observe, that the eigenmode occupations are rearranged due to a different mode transformation ($\nGP\neq \texttt{const}$) and minor coherences are introduced by the convolution.
}
\label{fig:smearing}
\end{figure}

\section{Extended data}
\label{sec:app:extendeddata}
Here we give further results on additional experimental scans that were performed in the study of revivals in Ref.~\cite{Recurrence}.
In total we consider 5 systems (one of them already presented in the main text) with varying system size and particle number --- the corresponding values are listed in Tab.~\ref{table_scans}.
\begin{table}[htb]
\begin{tabular}{c | c | c }
Scan &  System size $L = 2 \RD ~$  & Average particle number per well $N^{\rm Avg}_{\rm Well}$  \\ 
\hline
1          &   $\SI{49 }{\micro\meter}$ & 3147.5\\
2          &   $\SI{60 }{\micro\meter}$ & 3813.6\\
3          &   $\SI{38 }{\micro\meter}$ & 2293.5\\
4          &   $\SI{43 }{\micro\meter}$ & 2625.7\\
5          &   $\SI{54 }{\micro\meter}$ & 3513.3\\
\end{tabular}
\caption{Table of the experimental scans performed characterized by the system size specified via the external trap and the average particle number per well. 
The fluctuations of the particle number is about  $\delta N^{\rm Avg}_{\rm Well} \approx 50$ independent of the system size.
These two parameters together with the harmonic longitudinal  $\omega_l = 2\pi \times \SI{7}{\hertz}$ and radial $\omega_\perp = 2 \pi \times\SI{1400}{\hertz}$ trapping frequencies allows to calculate the Gross-Pitaevskii profile $\nGP$ and hence parametrize the effective model.
The first scan is the one presented in the main text.}
\label{table_scans}
\end{table}

In Fig.~\ref{fig:ini_reconstructions_many} we show the reconstructed covariance matrices of the initial state as described in the main text for all 5 experimental scans listed in Tab.~\ref{table_scans}.
Consistently with the result presented in the main text, we find that the reconstructed covariance matrices are close to being diagonal with significant squeezing suppressing phase fluctuations and enhancing density fluctuations.
Furthermore, in Fig.~\ref{fig:ini_reconstructions_many_conv} we show the covariance matrix obtained for the first scan considering wave functions convoluted with a Gaussian distribution \eqref{eq:SM_f_conv}.
Here we find similar $V_{j,k}$ to the covariance matrix shown in the main text for low lying modes but it is noticeable that the higher modes are populated with no clear decay tendency.
This can be explained by overfitting noise as the convolution of wave functions for large $k$ vanishes $\fjp(z)\approx 0$.
Indeed, examining \eqref{eq:SM_ls} we find that the optimizer is not sensitive to changes of $V_{j,k}$ with $j,k$ above the effective cut-off, or in other words the least squares recovery becomes numerically ill-conditioned.
We did not observe any significant improvement in quantitatively predicting the revivals using convoluted modes.

Secondly, we investigate in Figs.~\ref{fig:revivals23} and \ref{fig:revivals45} the correlator $C$ defined in the main text based on data obtained in scans $2$ to $5$. 
We show the values extracted from the experimental measurement as well as the results obtained from three reconstructions with different input intervals. 
The results are consistent with the ones presented and discussed in the main text. The experimental data shows a slow dephasing and weakening of the initial phase locking. 
The reconstructions are able to recover and predict the signal well if the reconstruction interval includes a recurrence. 
Reconstructions from dephased data (reconstruction regions a) and c)\,) are able to qualitatively describe the system but fail to predict quantitatively for instance the strength of the recurrence.

\begin{figure}[H]
\centering

\includegraphics[ width=0.99\linewidth]{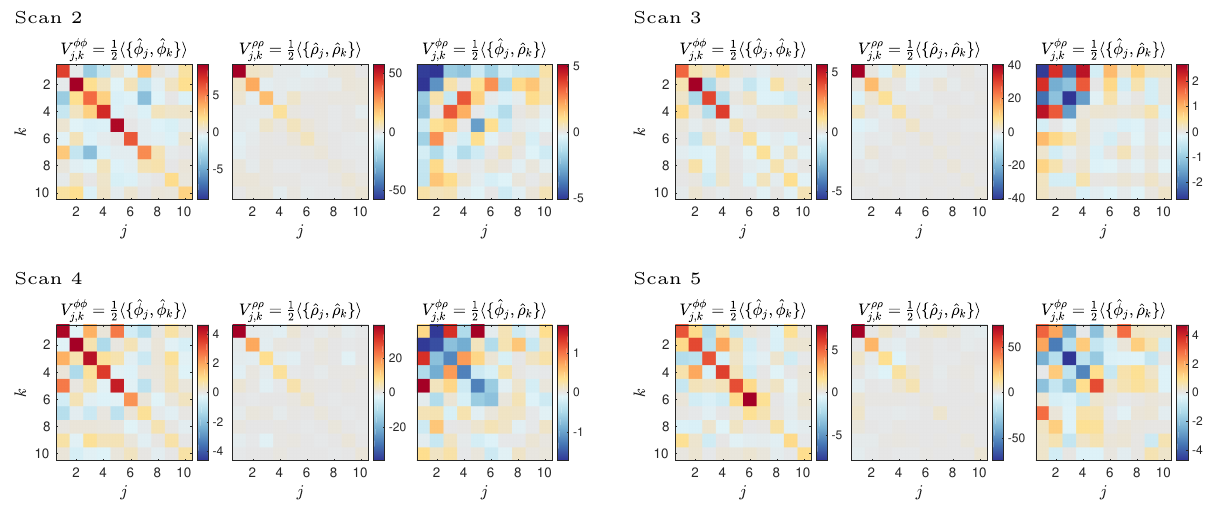}
\caption{We show as in Fig.~2 in the main text the blocks of the covariance matrix of the initial state reconstructed with $\N = 10$ modes for all experimental scans. The system sizes and particle number per well are as given in Table~\ref{table_scans}.}
\label{fig:ini_reconstructions_many}
\end{figure}
\begin{figure}[H]
\center
\includegraphics[trim = 0cm 8.1cm 0cm 11.4cm, clip,  width=0.49\linewidth]{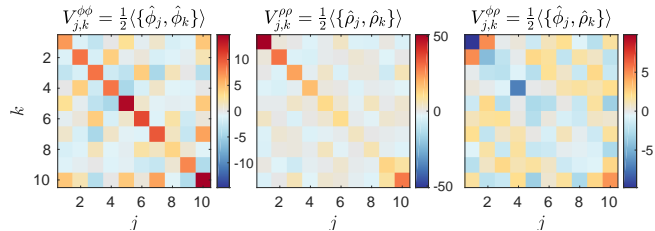}
\caption{We show as in Fig.~\ref{fig:ini_reconstructions_many} the reconstructed covariance matrix of the initial state of the first scan, but using convoluted eigenfunctions.
As modes with increasing energy display more and more oscillations, the convoluted modes with higher energy become smaller in amplitude once the convolution starts to average over a full oscillation. This renders the least squares less stable and higher modes can have large occupation numbers without changing the real-space correlations because the mode functions are suppressed by the convolution.
}
\label{fig:ini_reconstructions_many_conv}
\end{figure}
\begin{figure}[H]
\centering
\begin{minipage}{0.45\textwidth}
\includegraphics[trim = 1cm 0.2cm 2cm 20cm, clip,width=0.95\linewidth]{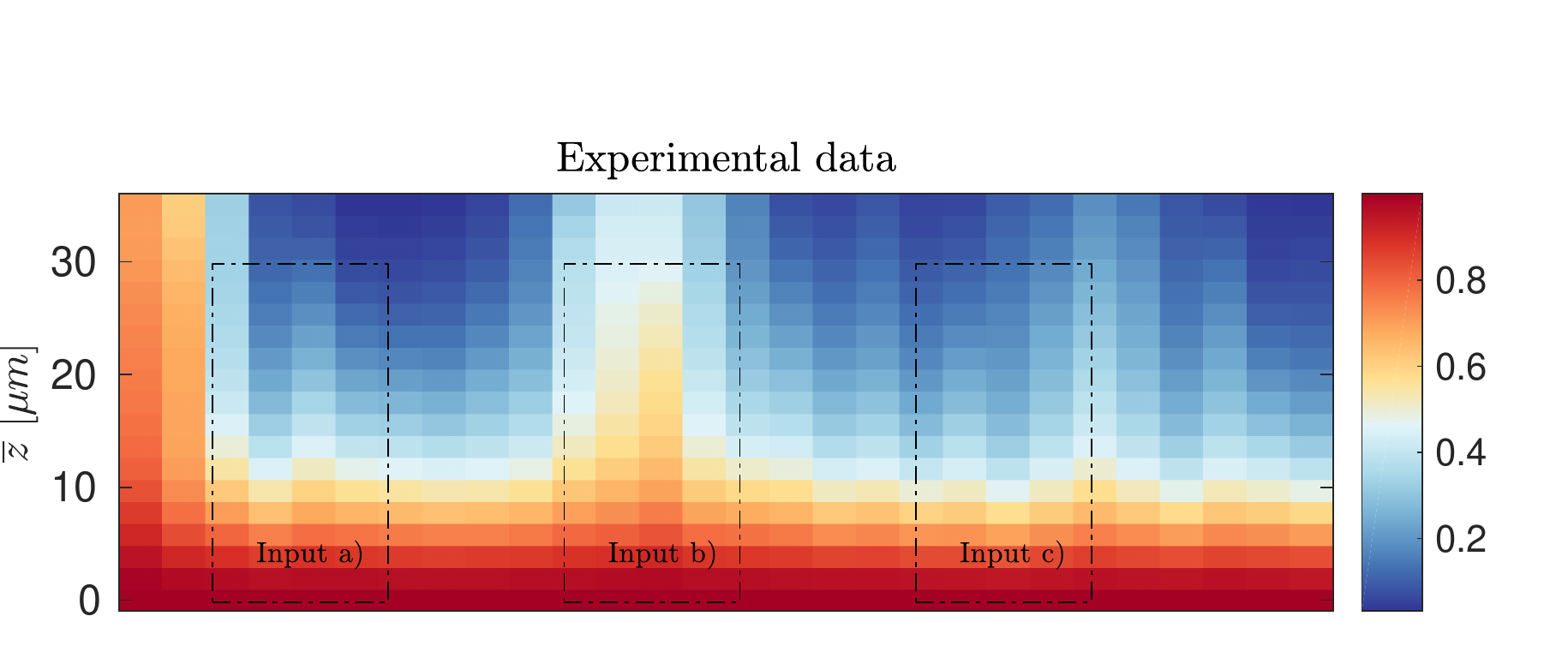}\\
\includegraphics[trim = 1cm 0.2cm 2cm 20.5cm, clip,width=0.95\linewidth]{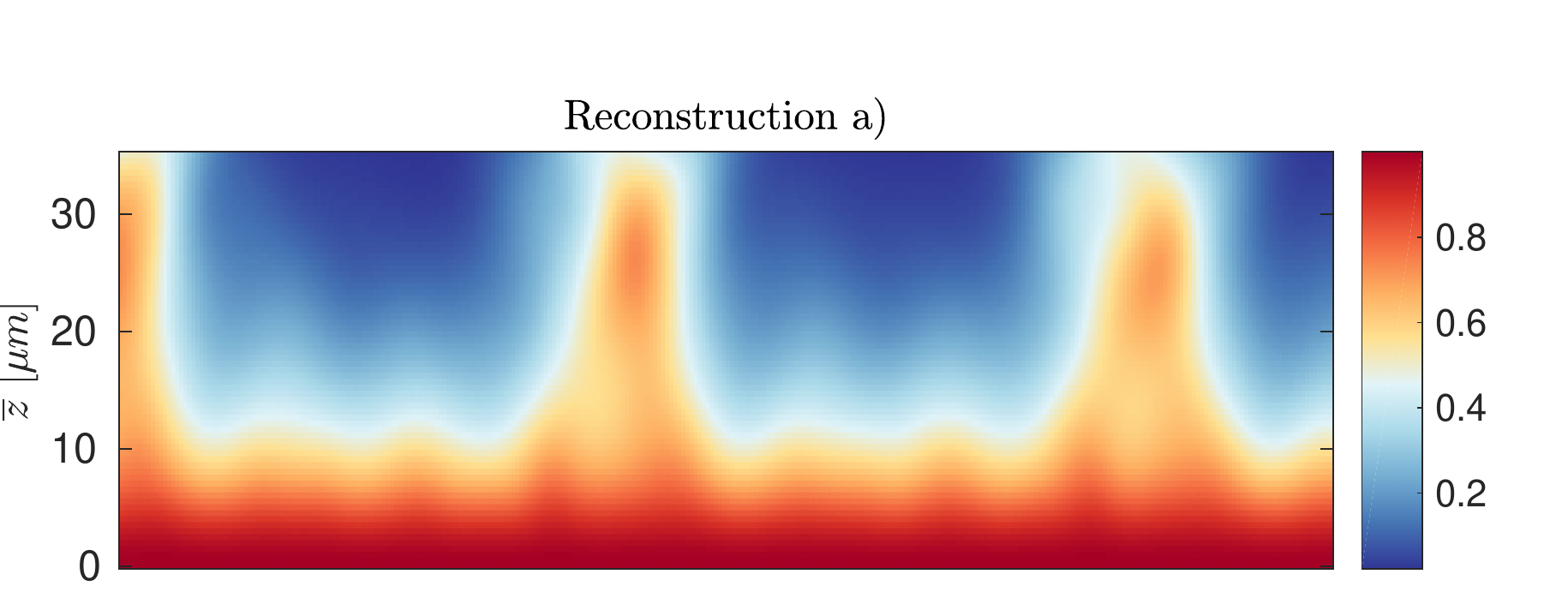}\\
\includegraphics[trim = 1cm 0.2cm 2cm 21.5cm, clip,width=0.95\linewidth]{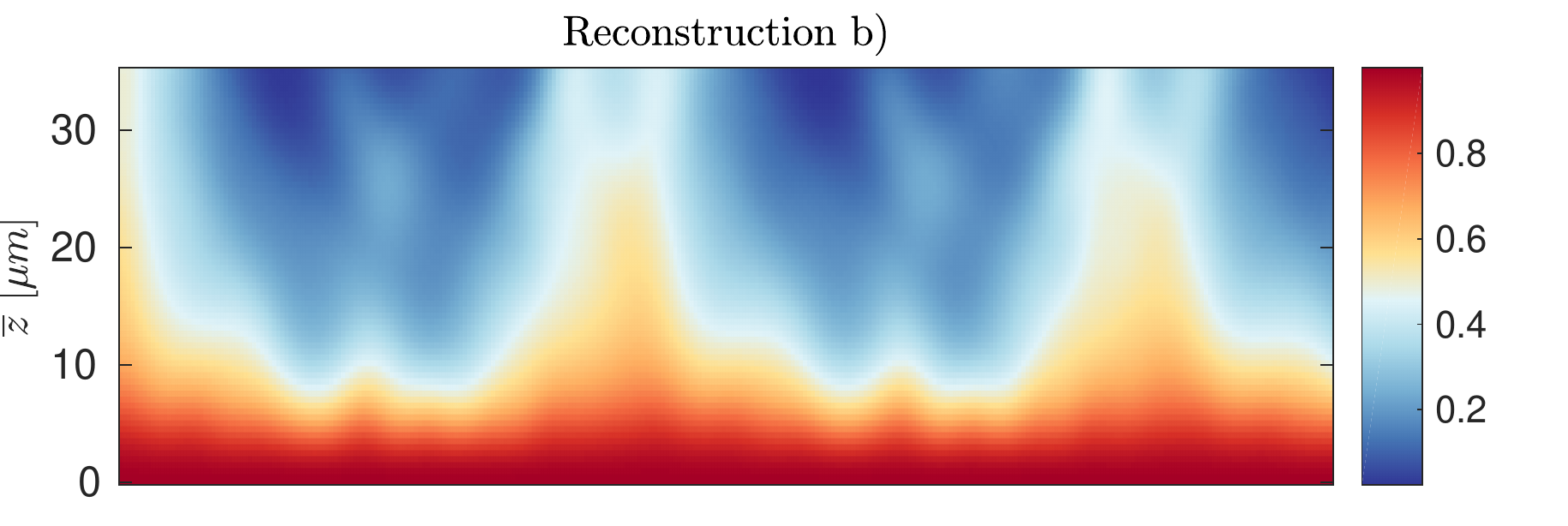}\\
\includegraphics[trim = 1cm 0.cm 2cm 20.8cm, clip,width=0.95\linewidth]{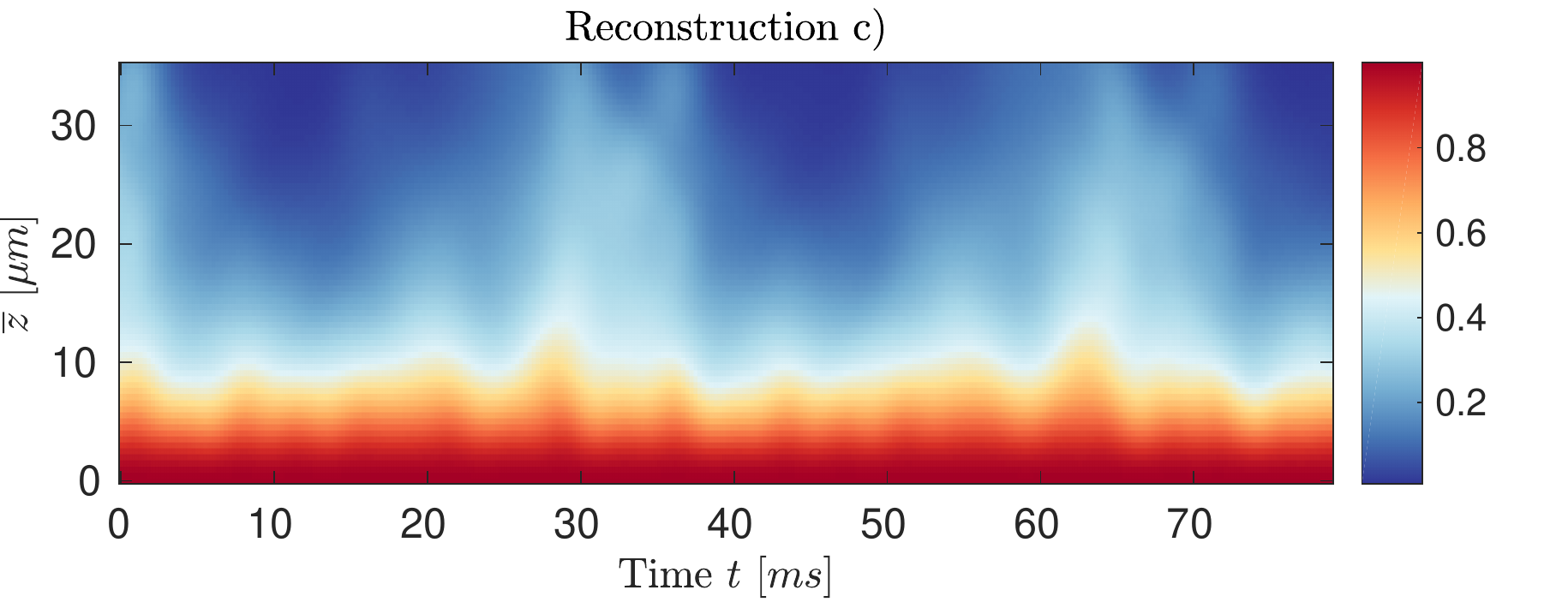}
\end{minipage}
\begin{minipage}{0.45\textwidth}
\includegraphics[trim = 1cm 0.2cm 2cm 20cm, clip,width=0.95\linewidth]{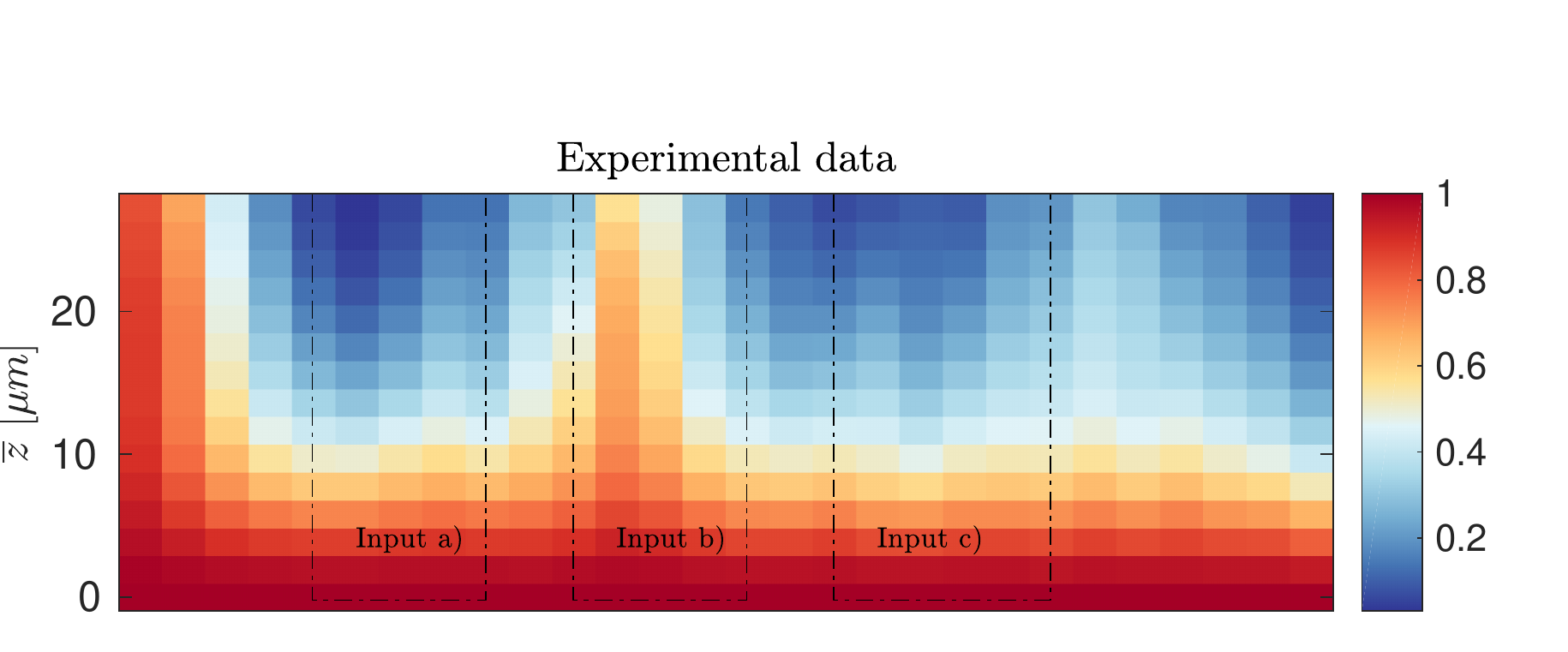}\\
\includegraphics[trim = 1cm 0.2cm 2cm 20.5cm, clip,width=0.95\linewidth]{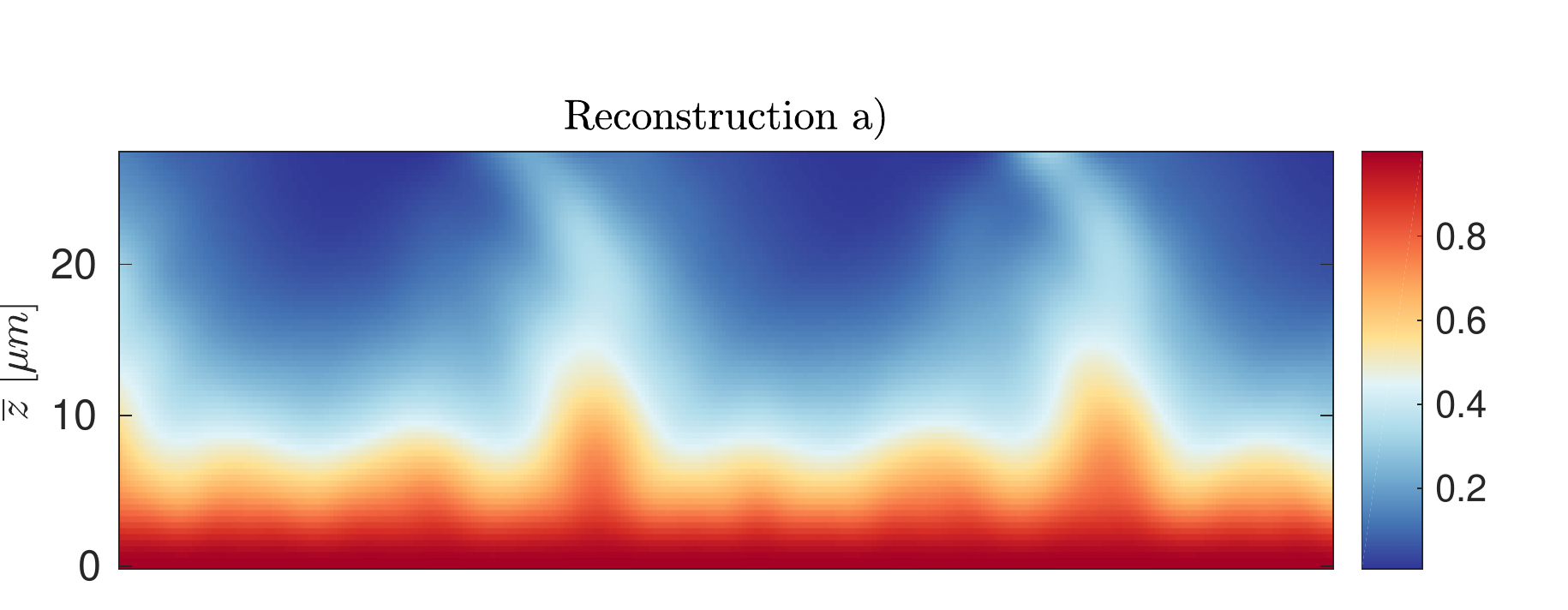}\\
\includegraphics[trim = 1cm 0.2cm 2cm 21.5cm, clip,width=0.95\linewidth]{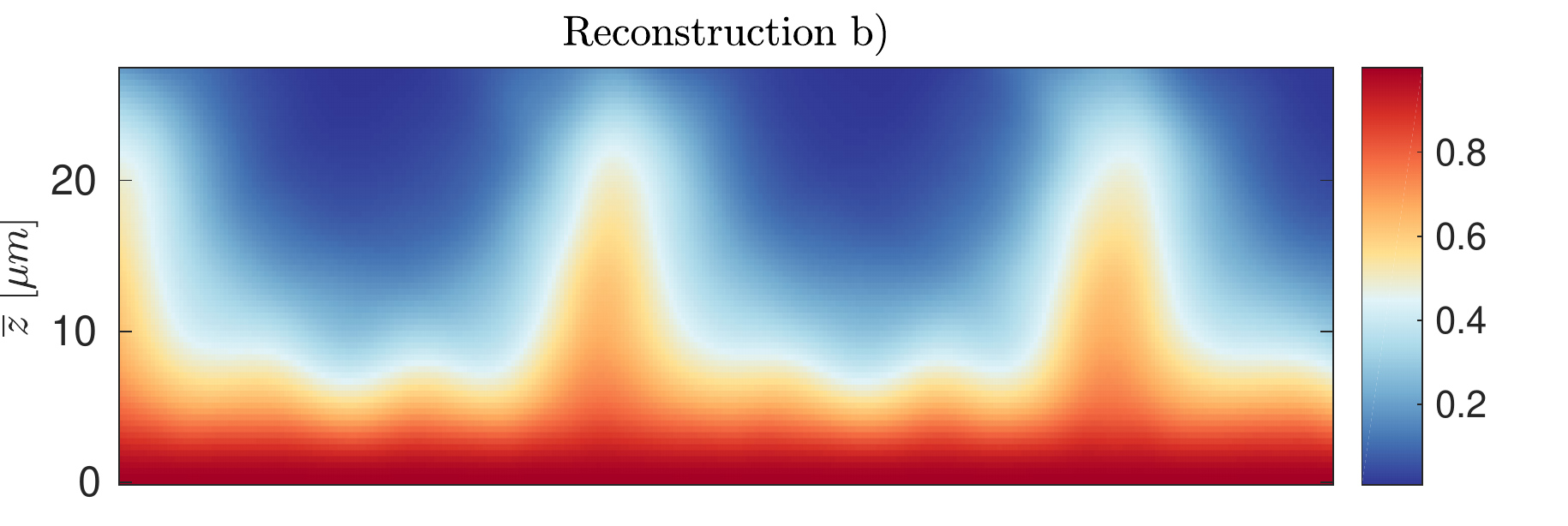}\\
\includegraphics[trim = 1cm 0.cm 2cm 20.8cm, clip,width=0.95\linewidth]{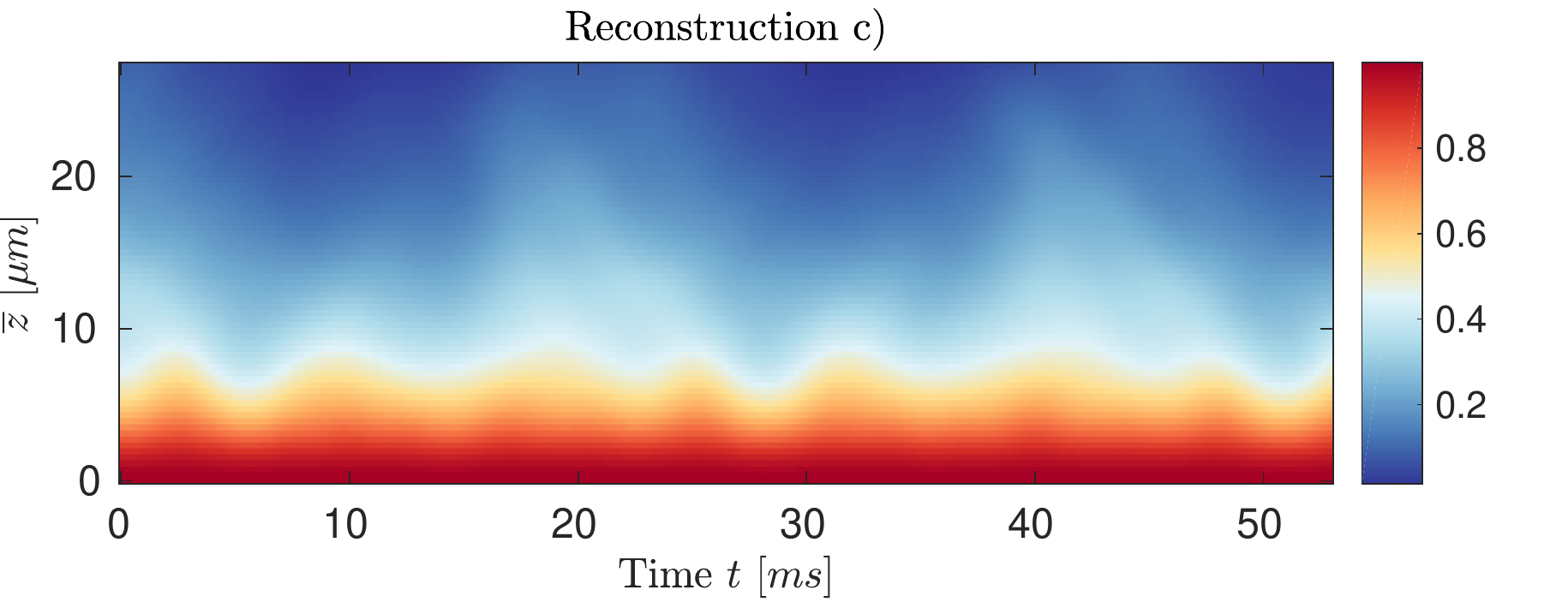}
\end{minipage}
\caption{Analogous to Fig.~3 from the main text, we show the correlator $C$ calculated from the experimental data and based on three reconstructions with varying input windows (indicated with dashed boxes) for scan number 2 (left) of the largest system size and number 3 (right) of the smallest system size.
For the  scan number 2 we have moved input window a) to an earlier time which results in very accurate reconstruction of the revival which would not be the case after moving the input window to a later time by one unit $\Delta t$.
Note, however, that the extrapolation works well which indicates that our method given enough input can yield very good results even with relatively small values of the  dynamical phase.
For the scan number 3 we can reconstruct reliably in the regions between the revivals.
Note that in both cases input window c) does not yield strong reconstructed revivals but they are timed well and also  in the experimental data the second revivals are not pronounced.
}
\label{fig:revivals23}
\end{figure}
\begin{figure}[H]
\centering
\begin{minipage}{0.45\textwidth}
\includegraphics[trim = 1cm 0.2cm 2cm 20cm, clip,width=0.95\linewidth]{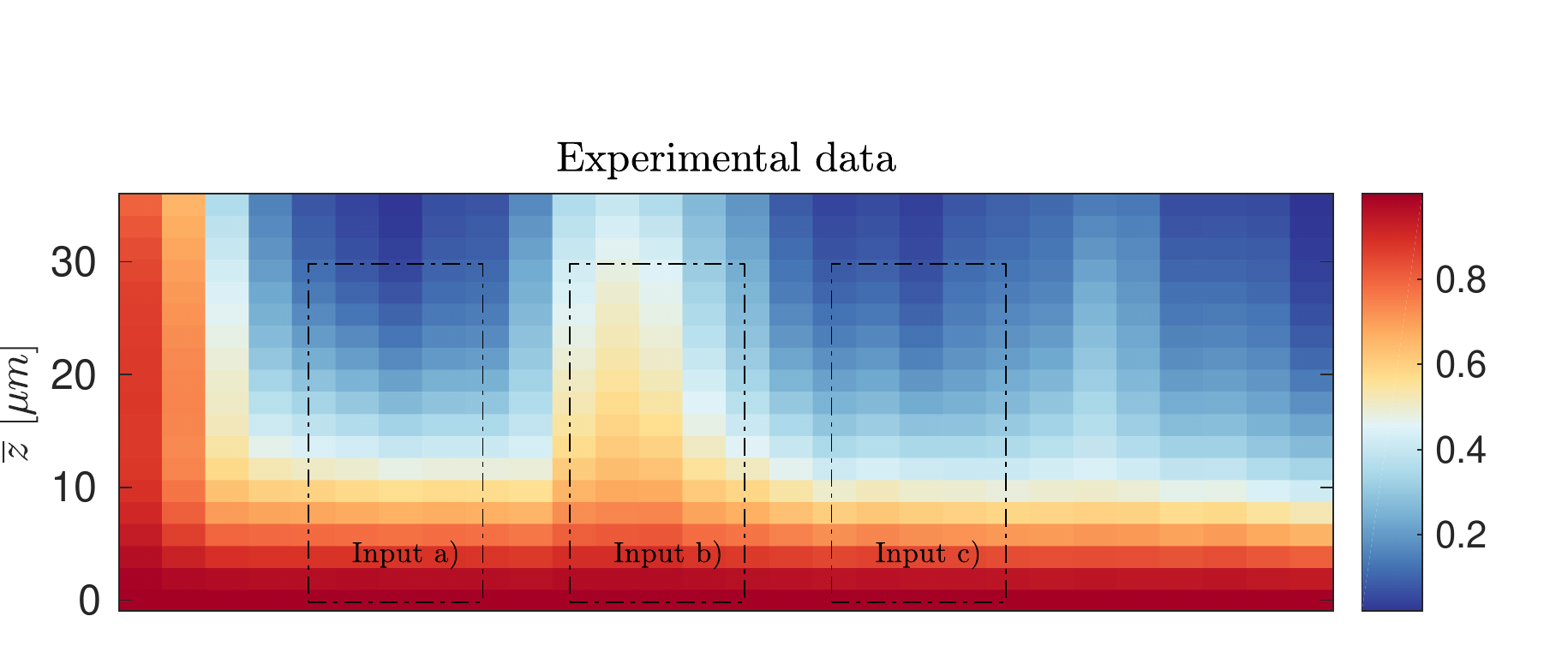}\\
\includegraphics[trim = 1cm 0.2cm 2cm 20.5cm, clip,width=0.95\linewidth]{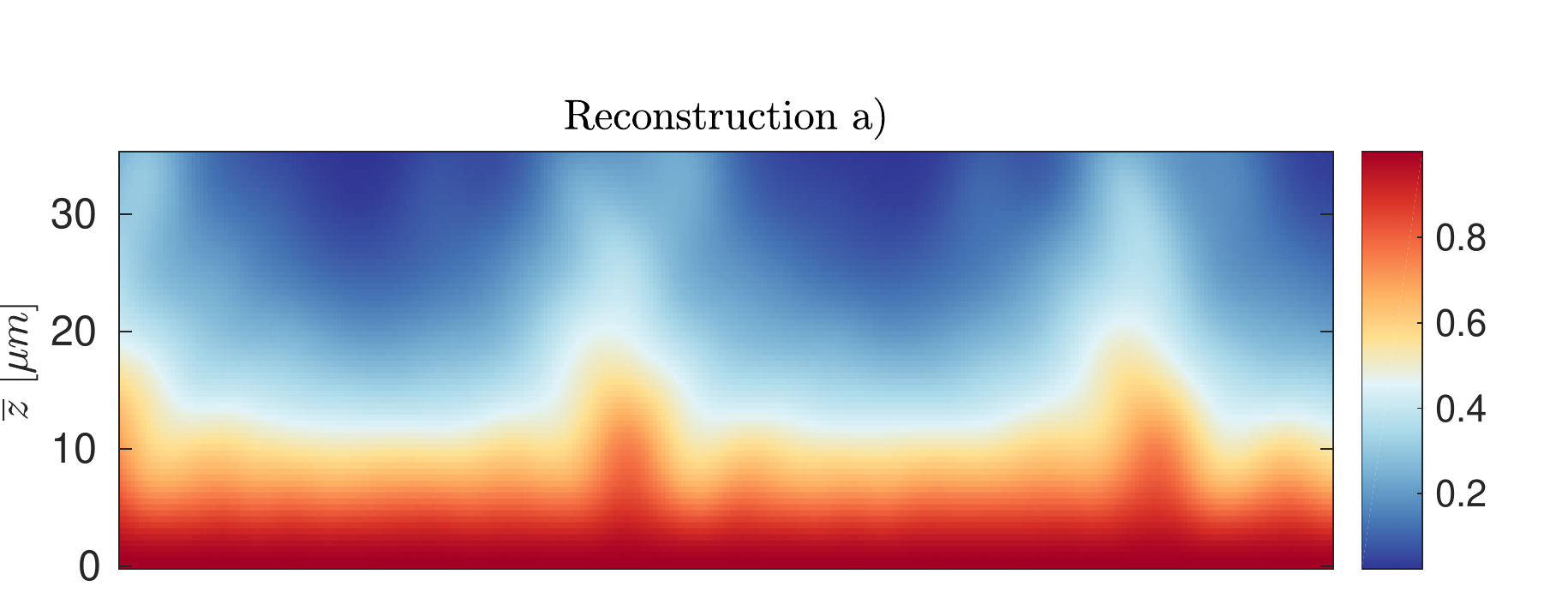}\\
\includegraphics[trim = 1cm 0.2cm 2cm 21.5cm, clip,width=0.95\linewidth]{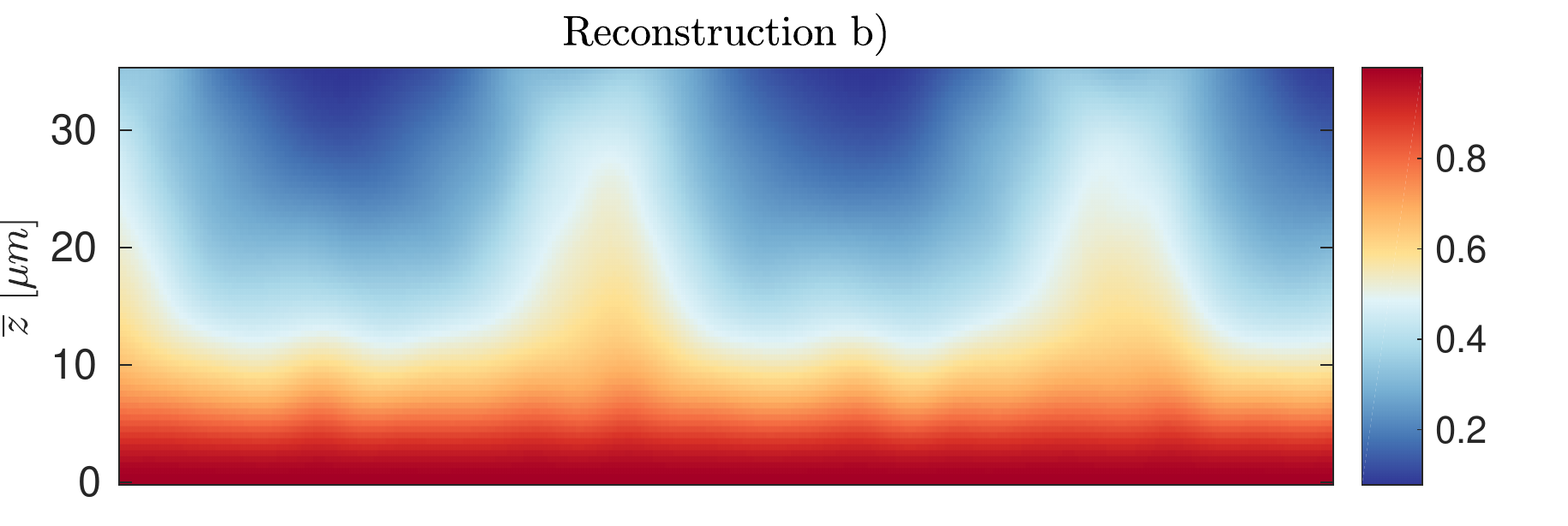}\\
\includegraphics[trim = 1cm 0.cm 2cm 20.8cm, clip,width=0.95\linewidth]{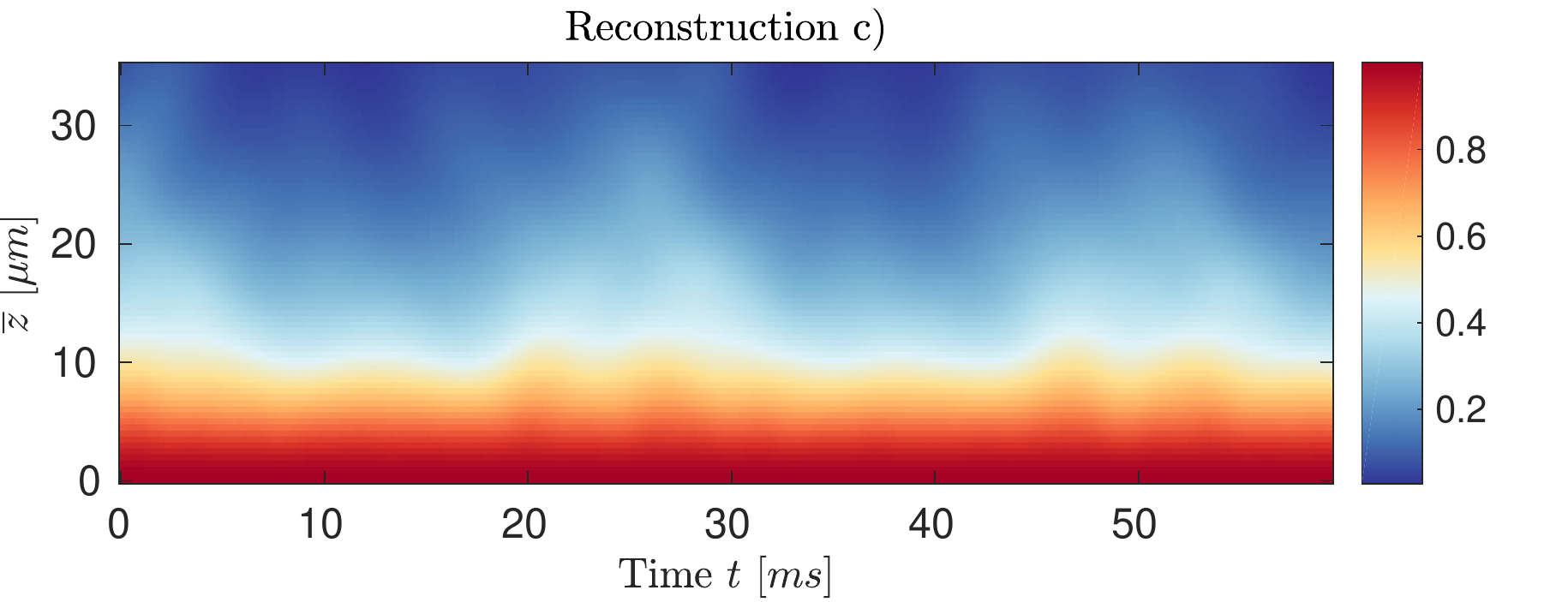}
\end{minipage}
\begin{minipage}{0.45\textwidth}
\includegraphics[trim = 1cm 0.2cm 2cm 20cm, clip,width=0.95\linewidth]{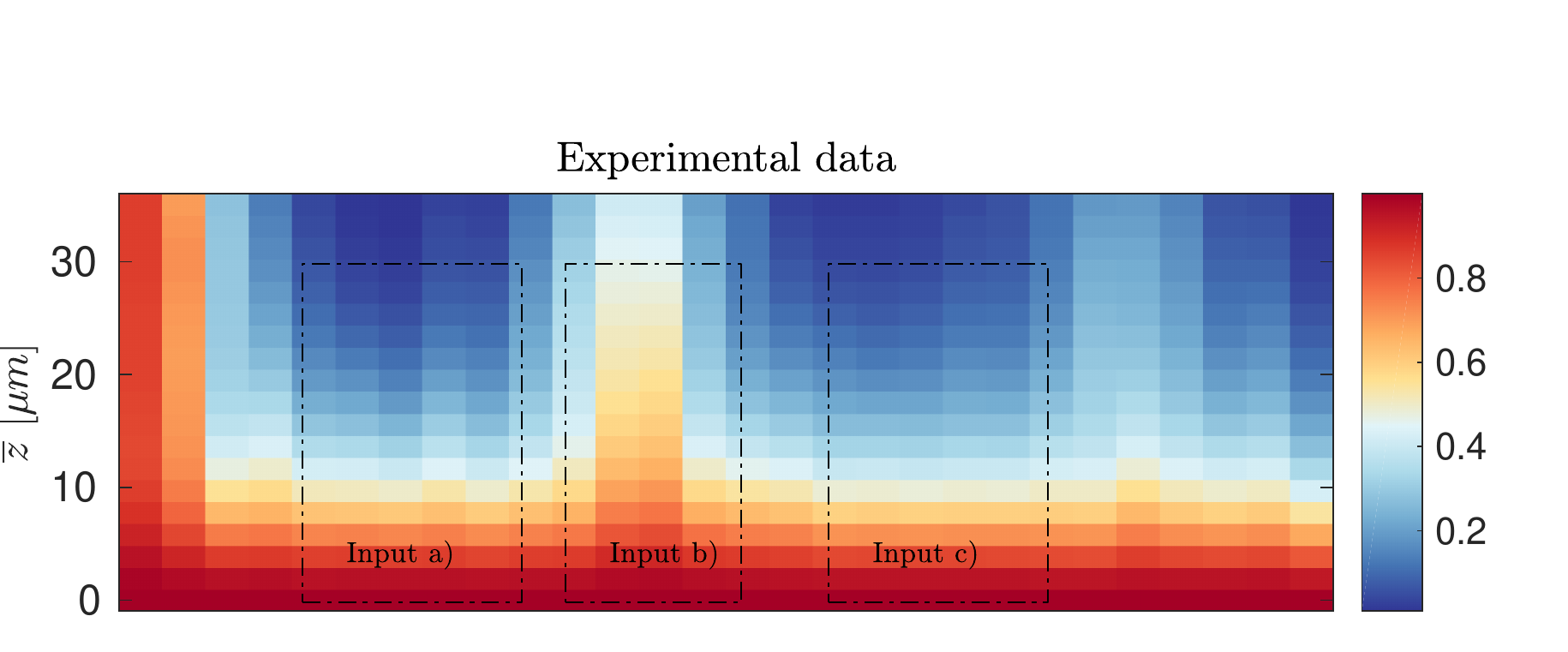}\\
\includegraphics[trim = 1cm 0.2cm 2cm 20.5cm, clip,width=0.95\linewidth]{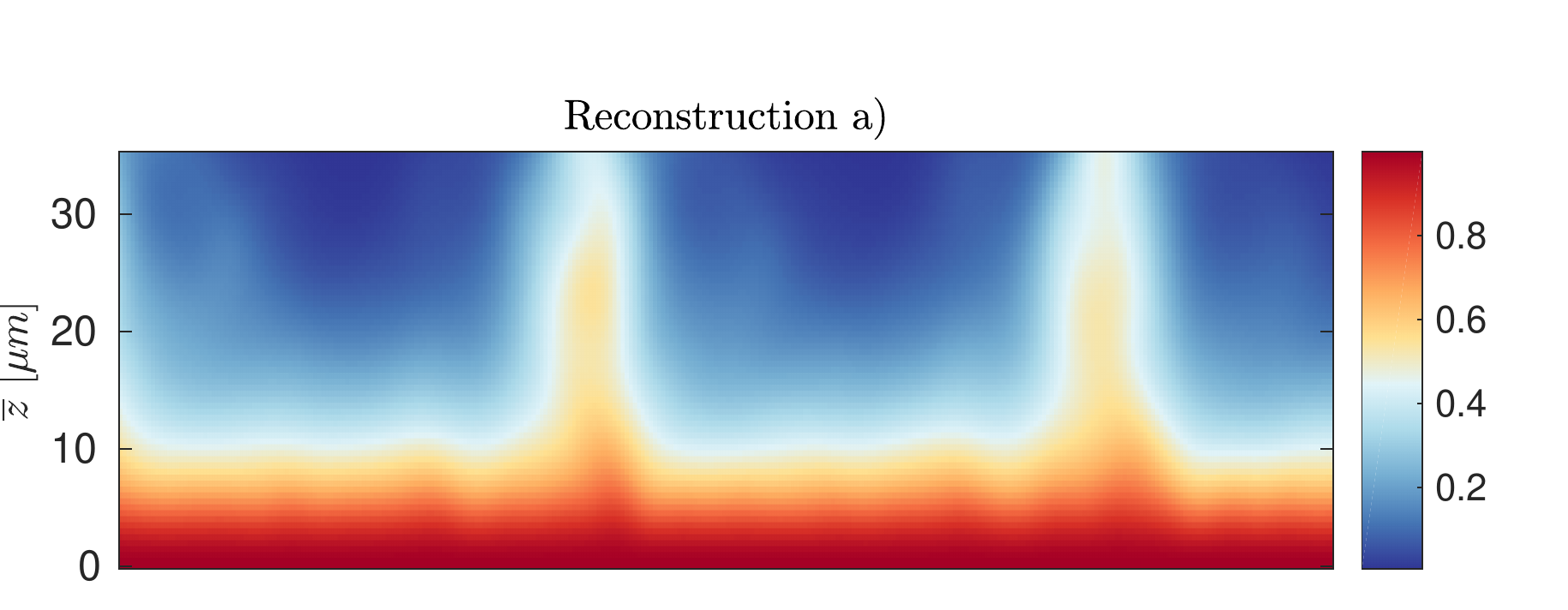}\\
\includegraphics[trim = 1cm 0.2cm 2cm 21.5cm, clip,width=0.95\linewidth]{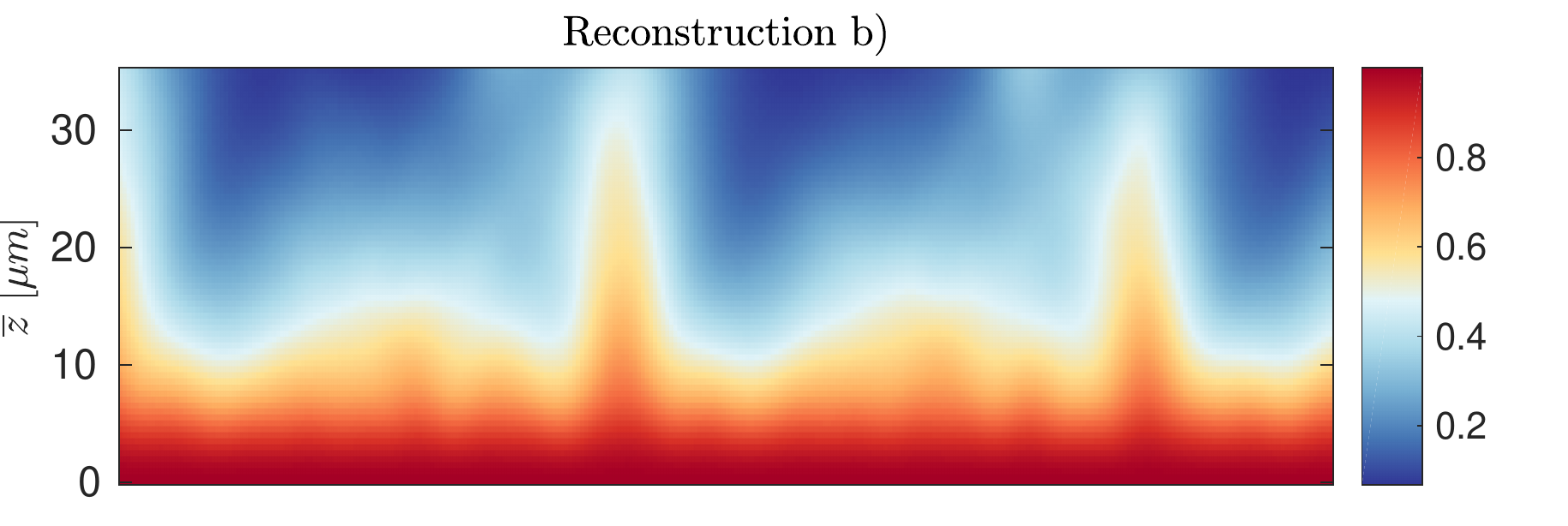}\\
\includegraphics[trim = 1cm 0.cm 2cm 20.8cm, clip,width=0.95\linewidth]{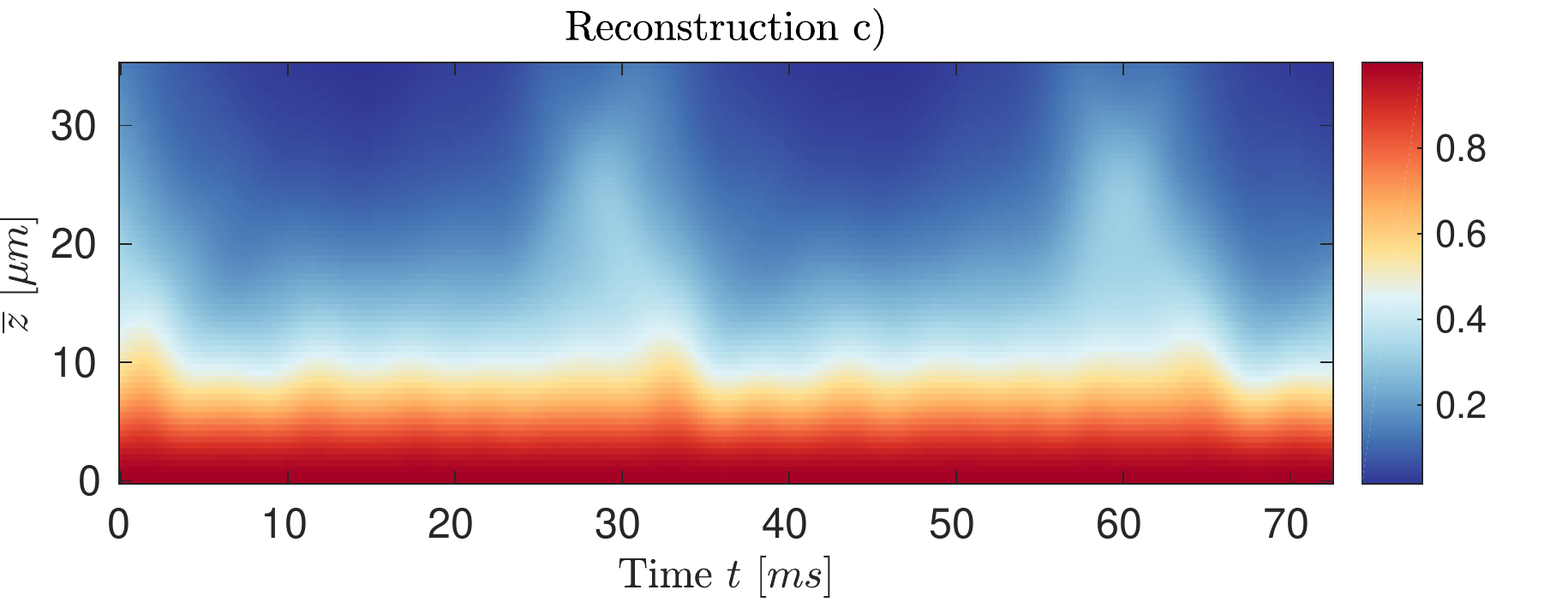}
\end{minipage}
\caption{As above for scan number 4 with second largest system size (left) and number 5 (right) with second smallest system size.}
\label{fig:revivals45}
\end{figure}

\begin{figure}[H]
\centering
\includegraphics[ width=0.99\linewidth ]{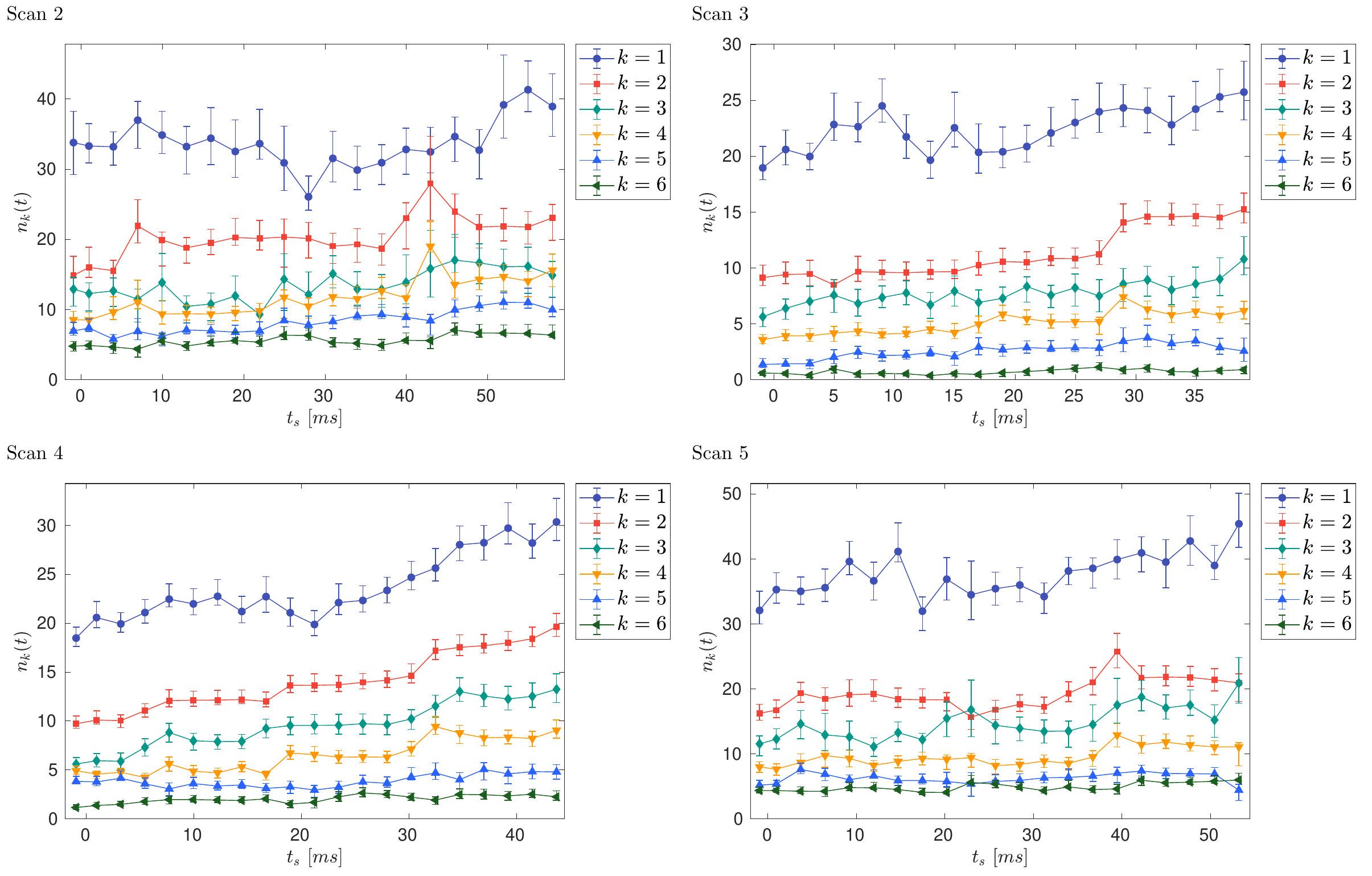}
\caption{Analogous to Fig.~4 in the main text we plot the occupation numbers for scans number 2-5 with parameters given in Table~\ref{table_scans} and bootstrap error-bars by resampling the phase profiles $n_{\rm Bootstrap}=500$ times. Note that the fluctuations of the reconstructed occupation of the first modes increase with increasing system sizes which shows that it is important that all the modes acquire enough dynamical phase.
Note that often the jumps in the occupation numbers coincide with the input intervals being placed in regions between the revivals where the reconstruction is difficult because of enhanced phase fluctuations due to the in-rotated density fluctuations.
We have checked that taking a larger number of input times $I$ does smoothen the occupation numbers but then the size of the input window is large enough so that interaction effects may start playing a role and the value of the occupation numbers need not be accurate. 
}
\label{fig:occupations25}
\end{figure}

\end{widetext}
\end{document}